\documentclass[prapplied,amssymb,amsmath,superscriptaddress,nofootinbib,twocolumn]{revtex4-2}
\usepackage{braket}
\usepackage{tikz}
\usepackage{multirow}
\usepackage{booktabs}
\usepackage[export]{adjustbox}
\usepackage{graphicx}
\usepackage{afterpage}
\usepackage[page,toc]{appendix}
\usepackage{amsthm}
\usepackage{lipsum}
\usepackage{dsfont}
\usepackage{hyperref}
\hypersetup{
	colorlinks=true,
	linkcolor=blue,
	filecolor=magenta,      
	urlcolor=cyan,
	citecolor=red
}
\usepackage{xcolor}
\usepackage[normalem]{ulem}
\usepackage{soul}
\AtBeginDocument{
	\heavyrulewidth=.08em
	\lightrulewidth=.05em
	\cmidrulewidth=.03em
	\belowrulesep=.65ex
	\belowbottomsep=0pt
	\aboverulesep=.4ex
	\abovetopsep=0pt
	\cmidrulesep=\doublerulesep
	\cmidrulekern=.5em
	\defaultaddspace=.5em
}
\DeclareMathOperator{\Tr}{Tr}

\usepackage{tabularx}

\newcounter{protocol}

\newcommand{\outprod}[1]{\ket{#1}\!\!\bra{#1}}
\newcommand{\proj}[2]{\ket{#1}\!\!\bra{#2}}
\usepackage{dcolumn}% Align table columns on decimal point
\usepackage{bm}
\usepackage{blkarray}
\usepackage{xparse}
\usepackage{mathtools}
\usepackage{subfiles}
\graphicspath{{./}{Figures/}}
\usepackage{xcolor}

\makeatletter
\newcommand\footnoteref[1]{\protected@xdef\@thefnmark{\ref{#1}}\@footnotemark}
\makeatother

\usepackage{float}

\begin{document}
	\preprint{APS/123-QED}
	%opening
	\title{Coherent State Assisted Entanglement Generation Between Quantum Memories}

    \author{Chaohan Cui}
    \thanks{These authors contributed equally.}
    \email{chaohan@umd.edu}
    \author{Prajit Dhara}
    \thanks{These authors contributed equally. Currently at RTX BBN Technologies.}
    \author{Saikat Guha}
    \affiliation{Department of Electrical and Computer Engineering, The University of Maryland, College Park, MD 20742}
    \affiliation{Wyant College of Optical Sciences, The University of Arizona, Tucson, AZ 85721}

    % \date{Started: March 26, 2024 \quad Last Updated: \today }
\begin{abstract}
Generating entanglement deterministically at a capacity-approaching rate is critical for next-generation quantum networks. We revisit previous protocols and present a readily-realizable coherent-state-assisted protocol that can generate entanglement near-deterministically between reflective-cavity-based quantum memories at up to twice the rate attainable using the dual-rail photonic-qubit mediated scheme. The most pronounced rate benefit occurs in the low-channel-loss regime and persists even with moderate noise. The protocol can be generalized to entangling an array of memories into various multi-qubit entangled states, including a GHZ state.
\end{abstract}
	
\maketitle
		%\tableofcontents

\section{Introduction}

The vision for the {\em quantum internet} is to enable quantum communications between a diverse group of users supporting various applications~\cite{kimble2008quantum,duan2010colloquium,wehner2018quantum,awschalom2021development,azuma2023quantum}. Central to this vision, is the need for quantum networks that can reliably and faithfully establish distributed entanglement among quantum memories~\cite{lvovsky2009optical}. Applications such as quantum-enhanced long-baseline telescopes~\cite{gottesman2012longer,khabiboulline2019optical,rajagopal2024towards,Padilla2024-in}, distributed quantum sensing~\cite{ge2018distributed,zhang2021distributed}, and distributed and blind quantum computing~\cite{jiang2007distributed,broadbent2009universal,barz2012demonstration,monroe2014large} are all expected to benefit from the development of these networks. Most currently-studied quantum network architectures are based on pairwise entanglement generation between two quantum memory banks separated by an optical channel, paired with entanglement distillation~\cite{wei2022towards,dhara2023entangling,azuma2023quantum}. Over the last decade, multiple experiments have successfully demonstrated photon-mediated entanglement generation between remote quantum memories over metropolitan (and ever-expanding) distances using various qubit-photon interfaces on platforms such as semiconductor quantum dots~\cite{stockill2017phase,you2022quantum,dusanowski2022optical,yu2023telecom}, trapped ions~\cite{moehring2007entanglement,stephenson2020high,krutyanskiy2023entanglement}, neutral atoms~\cite{ritter2012elementary,hofmann2012heralded,van2022entangling,liu2024creation}, and solid-state defects~\cite{bernien2013heralded,pompili2021realization,knaut2024entanglement}. 

The aforementioned experimental demonstrations rely on the transmission and interference of single photons. The most common approaches begin with the generation of memory-qubit-photon entanglement, either by direct memory-state-dependent photon emission~\cite{ritter2012elementary,hofmann2012heralded,bernien2013heralded,delteil2016generation} or by imparting a qubit-dependent phase on a single photon wavepacket~\cite{duan2004scalable,sipahigil2016integrated}. The photons are transmitted from two remote quantum memories and are measured in the Bell basis by a partial Bell state measurement realized by optical interference at a central site (often called a {\em midpoint swap} quantum link). Alternatively, one party acts as a transmitter and the other as the receiver, where the transmitted photon---entangled with the transmitter-side memory---interacts with and transfers its quantum state into the receiver-side memory~\cite{ritter2012elementary,bhaskar2020experimental}. Depending upon the photonic encoding of the qubit, these {\em single-photon protocols} may either be classified as single- or dual-rail~\cite{furusawa2011quantum}. The single-rail protocol post-selects on the entanglement-swap outcome corresponding to the detection of a single photon, whereas the dual-rail protocol post-selects a successful swap based on a two-photon (i.e., Hong-Ou-Mandel~\cite{hong1987measurement}) detection outcome. For midpoint swap links over an optical channel of transmissivity $\eta$, the single-rail protocol achieves a better rate-loss scaling of its entanglement generation rate $\mathcal{R} \propto \sqrt{\eta}$ in the high-loss (long-distance) regime ($\eta \ll 1$). The dual rail protocol suffers from a worse rate-loss scaling of $\mathcal{R}\propto\eta$ when $\eta \ll 1$. However, for a fixed set of link and hardware parameters, the fidelity of the distributed entangled state (to an ideal Bell state) is generally higher for the dual-rail protocol. Detailed comparisons of these two single-photon protocols were presented in Refs.~\cite{dhara2023entangling,Hermans2023-kf}.
In the low-loss regime, both protocols saturate at a maximum $\mathcal{R}=0.5$ entangled qubit pairs generated per channel use (i.e., ebit/ch), which is far lower than the repeater-less bound for midpoint-swap schemes, given by $Q_2(\sqrt{\eta})=-\log_2(1-\sqrt{\eta})$~\cite{pirandola2009direct,pirandola2017fundamental}, especially when $\eta$ is high (low loss).

High-rate multi-qubit entanglement generation is crucial for many applications, such as preparing local ancillas for measurement-based quantum error correction for communications~\cite{munro2012quantum,slussarenko2022quantum,hilaire2023linear}, quantum repeaters for multi-site entanglement~\cite{briegel1998quantum,azuma2015all,pant2017rate}, and memory-assisted quantum switches or routers~\cite{pant2019routing,vardoyan2020exact,lee2022quantum}. This pairwise heralding success probability results in more attempts per entanglement generation, reducing the rate of generating multi-qubit entanglement. In addition to this inefficiency in the low-loss regime, single-photon protocols also suffer from susceptibility to imperfect distinguishability (a.k.a mode mismatch) between the temporal-spectral mode shapes of single photons produced by independent emitters, which poses further challenges to their scalability. 

\begin{figure*}[htbp]
\centering
\includegraphics[width=\linewidth]{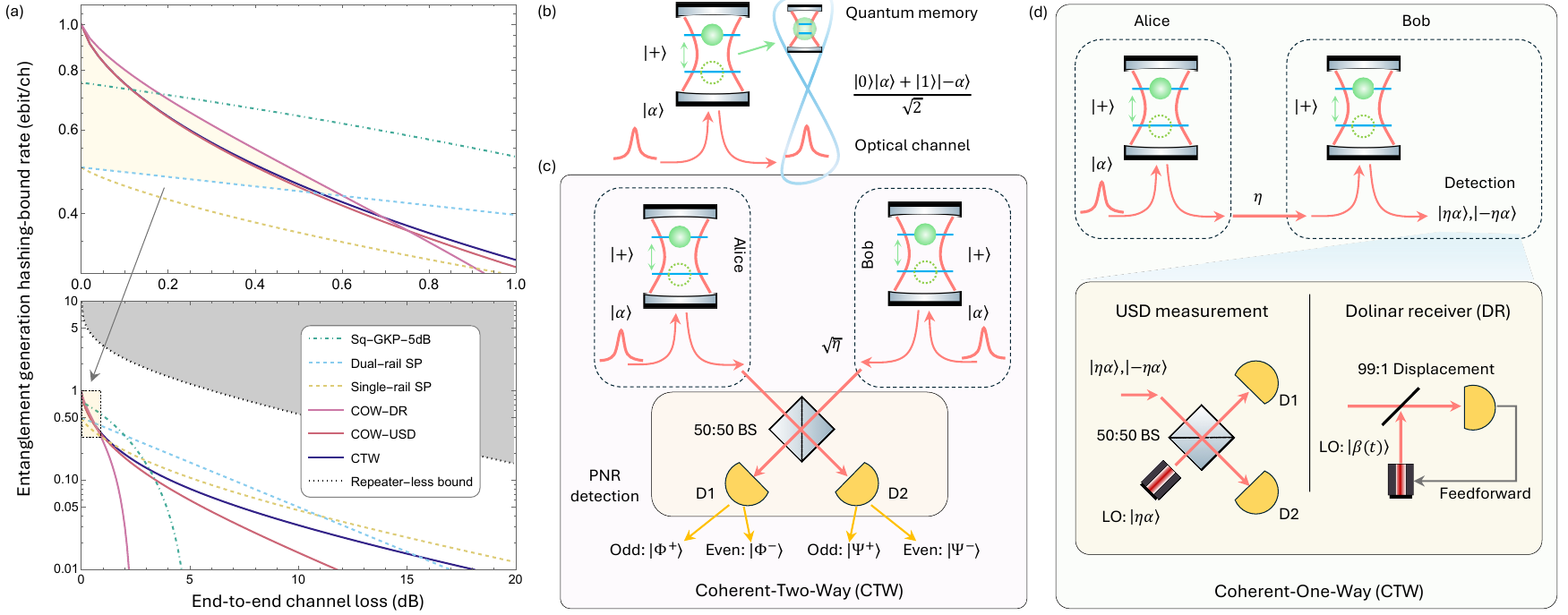}
\caption{(a)Evaluation of entanglement generation Hashing-bound rates of different protocols~\cite{dhara2024entangling,dhara2023entangling}. The advantageous area of coherent-state-assisted protocols is highlighted. (b) Control-phase gate between the memory qubit and the coherent state realized by the reflective-cavity-based qubit-photon interface. (c) CTW entanglement generation protocol. (d) COW entanglement generation protocols with two different measurement strategies: unambiguous state discrimination (USD) and Dolinar receiver (DR).}
\label{fig:Fig123} 
\end{figure*} 

Recently, several continuous-variable solutions have been proposed to boost the entanglement rate towards the repeater-less bound by using optical Gottesman-Kitaev-Preskill (GKP) qudit encoding~\cite{dhara2024entangling} or via quantum nondemolition (QND) measurements in lieu of single photons and projective measurements~\cite{winnel2022achieving}. However, near-term limiting factors for realizing these protocols are the practical quality and robustness of GKP state generation~\cite{konno2024logical} and the practical feasibility of the photonic QND measurements.
In this article, we revisit and propose three categories of protocols for entangling quantum memories utilizing coherent-state pulses, and on-off single-photon detector or photon-number-resolving (PNR) detection. We analyze the rates achieved by all the protocols and show that they can outperform the two well-known single-photon protocols and the recent 5 dB GKP system in the low-loss regime [Fig.1(a)]. All three categories of protocols can be generalized for distributing a multi-qubit state among an array of quantum memories. Besides, preparing a coherent state in a desired temporal-spectral mode prior to the qubit-photon interface is far easier, more compatible with frequency conversion~\cite{bersin2024telecom}, and more scalable than using single-photon emitters. These practical benefits make the proposed protocols more feasible in the near term. We describe our protocols and highlight their achievable distillable entanglement rate per channel use by deriving the Hashing-based lower bound~\cite{devetak2005distillation}. We discuss the impact of imperfections in the memory-qubit-photon interface, interference mode mismatch, and imbalances in the post-memory-reflection coherent-state amplitudes on the quality and rate of entangling multiple memories, posting quantitative benchmarks for device-level performance targets.

% \begin{figure}[htbp]
% \centering
% \includegraphics[width=0.7\linewidth]{Fig1_v1.pdf}
% \caption{(a) Control-phase gate between memory qubit and the coherent state realized by reflective-cavity-based qubit-photon interface. (b) CTW entanglement generation protocol.}
% \label{fig:Fig1} 
% \end{figure} 

\section{Protocols}
Central to the protocols proposed for coherent-state-assisted entanglement generation are quantum memories that interact with an optical mode by a reflective-cavity-based memory-qubit-photon interface. As shown in Fig.~\ref{fig:Fig123}(b), this type of interface imparts an optical $\pi$-phase shift to the reflected optical mode, conditioned on the qubit state of the quantum memory, which has been theoretically investigated~\cite{duan2004scalable,raymer2024duan,reiserer2015cavity,Dhara2024-ko,waks2006dipole} and experimentally realized in multiple platforms~\cite{wang2005engineering,sun2016quantum,sipahigil2016integrated,daiss2019single,hacker2019deterministic,vasenin2024evolution}. Depending on how the coherent states are encoded, transmitted, interfered, and detected between two quantum memories, we categorize the protocols discussed in context into three sets: coherent-two-way (CTW), coherent-one-way with unambiguous state discrimination (COW-USD), and coherent-one-way with Dolinar Receiver (COW-DR) to practically achieve the minimum error of discriminating two coherent states. We will show how these three categories can be generalized to entangling more than two quantum memories.

Let us begin with the coherent-two-way (CTW) protocol and show how this coherent-state-assist protocol could beat the single-photon protocols. In this section, we assume the memory-qubit-photon interfaces are ideal, i.e., without noise and imperfections. Considering the scheme depicted in Fig.~\ref{fig:Fig123}(c), the two quantum memories ready for photon-mediated entanglement generation are located at the two sites, named Alice and Bob. Both sites first prepare the qubit state of their own quantum memory to $|+\rangle$ state and interact their memory qubit with a locally-generated coherent state $|\alpha\rangle$ in an optical mode. After the reflective-cavity-based interaction, the memory-qubit state is entangled with the phase of the outgoing coherent state. 
\begin{equation}
|\psi\rangle_{k}=\frac{|0\rangle_{k}|\alpha\rangle_{k}+|1\rangle_{k}|-\alpha\rangle_{k}}{\sqrt{2}}; k=\{A,B\}
\end{equation}
For brevity, the subscript $k$ labels both the memory qubit and its associated optical mode. Both parties then transmit their optical mode over a pure-loss optical channel of equal transmissivity $\sqrt{\eta}$ toward a central site with an apparatus to perform the entanglement swapping. 

After transmission through each arm of intensity transmissivity $t=\sqrt\eta$, the joint state of the memory, optical mode, and environment mode $E_k$ is
    \begin{align}
        \ket{\psi'}_{kE}=\frac{1}{\sqrt{2}}\biggl( &\ket{0}_k \ket{\alpha\sqrt{t}}_k \ket{\alpha\sqrt{1-t}}_{E_k} \\
        &+\ket{1}_k \ket{-\alpha\sqrt{t}}_k \ket{-\alpha\sqrt{1-t}}_{E_k}\biggr) 
    \end{align}
    Tracing out the environment mode yields the state
    \begin{align}
        \begin{split}
        \rho_{k} = \frac{1}{2}\sum_{i,j=0}^{1} &\Theta_{ij}(\alpha;t) \proj{i}{j}_k \\
        &\otimes\proj{(-1)^i\alpha\sqrt{t}}{(-1)^j\alpha\sqrt{t}}_k
        \end{split}
    \end{align}
    where we define the environment overlap
    \begin{align}
    \Theta_{ij}(\alpha;t) = \braket{(-1)^j\alpha\sqrt{1-t}|(-1)^i\alpha\sqrt{1-t}}.
    \end{align} The term $\Theta_{ij}(\alpha;t)$ accounts for the photon loss induced dephasing incurred by the joint state, which is the main source of error for the coherent-state-assisted entanglement distribution protocols featured in this article.

Then, the entanglement swapping measurement will be applied on the optical modes of $\rho_A$ and $\rho_B$. The measurement apparatus comprises a 50:50 beamsplitter (for the interference of the memory-qubit-conditioned weak coherent states) followed by one PNR detector at each output port (labeled $D1$ and $D2$). The state after interference (with the undetected environment modes) is
\begin{align}
    \ket{\varphi}_{AB} = \frac{1}{2}\sum_{i,j} \ket{i,j}_{AB} \ket{\beta_{i,j} (\alpha,\eta)}_{AB} 
    \ket{\gamma_{i,j} (\alpha;\eta)}_E
\end{align}
where we define
\begin{align}
    \Ket{\beta_{i,j} (\alpha;\eta)} &= \begin{cases}
        \ket{(-1)^{i} \alpha\sqrt{2\sqrt{\eta}},0} ; i=j\\
        \ket{0,(-1)^{i} \alpha\sqrt{2\sqrt{\eta}}} ; i\neq j\\
    \end{cases}\\
    \ket{\gamma_{i,j} (\alpha;\eta)}_E &\equiv \Ket{(-1)^i \alpha\sqrt{1-\sqrt{\eta}},(-1)^j \alpha\sqrt{1-\sqrt{\eta}}}_E
\end{align}

The measurement works similarly to what has been proposed and thoroughly tested in the twin-field quantum key distribution systems~\cite{lucamarini2018overcoming,wang2019beating}. The click pattern of the PNR detectors heralds the generation of entanglement between the quantum memories -- the pattern can be separated into five possible cases: (1) odd number photons detected at $D1$; (2) even number photons detected at $D1$; (3) odd number photons detected at $D2$; (4) even number photons detected at $D2$; (5) no photon detected at either detector. 

The first four outcomes are the {\em success} outcomes as they herald the generation of entanglement between the quantum memories -- the corresponding Bell states are shown at the bottom of Fig.~\ref{fig:Fig123}(c). In addition to reducing the success chance of heralding the entanglement generation, the channel loss also induces phase-flip errors in the heralded entangled states, thereby reducing the entangled state's fidelity and distillable entanglement. The lower bound to the achievable distillable entanglement generation rate, $\mathcal{R}_{\rm CTW}$, can be derived by evaluating the product of the total heralding probability and Hashing bound per heralded success (see Appendix A),
\begin{align}
    \mathcal{R}_{\rm CTW}= 
\left(1-e^{-2\sqrt{\eta}|\alpha|^2}\right)\left[1-h_2\left(\frac{1+e^{-4(1-\sqrt{\eta})|\alpha|^2}}{2}\right)\right]
\end{align}
The factor $1-e^{-2\sqrt{\eta}|\alpha|^2}$ is the heralding success probability, corresponding to the detection of at least one photon. The detection port and photon-number parity jointly identify the target Bell state. Under the ideal-interface and pure-loss assumptions considered here, channel loss introduces only dephasing, so the heralded state is a mixture of the target Bell state and its phase-flipped counterpart, with phase-flip probability $p=[1-e^{-4(1-\sqrt{\eta})|\alpha|^2}]/2$. The hashing yield per successful heralding event is therefore $1-h_2(p)$, where $h_2(x)=-x\log_2 x-(1-x)\log_2(1-x)$ is the binary entropy function. Multiplying this yield by the heralding success probability gives the rate above.

For a given $\eta$, there is an optimal initial coherent state amplitude $\alpha$ that maximizes $\mathcal{R}_{\rm CTW}$. A larger amplitude $\alpha$ increases the success probability but also increases the information leakage to the environment (which degrades the Hashing bound). For a channel with very low loss, i.e., $\eta\rightarrow1$, the Hashing rate approaches 1 ebit/ch for the optimal $\alpha$, which means the entanglement generation becomes deterministic, 3 dB higher than the highest-reachable rate of single-photon protocols. This pronounced advantage comes from the fact that this coherent-state-assisted protocol leverages the phase information embedded in the higher-photon-number Fock basis of the coherent state. In the high-loss regime ($\eta\ll1$), using the CTW protocol leads to a slight disadvantage compared to single-photon protocols as these high-photon-number Fock basis components leak more information to the environment, causing more phase error that outweighs the improved success probability. In this regime, $\mathcal{R}_{\rm CTW}\approx 0.07\sqrt{\eta}$ versus the single-rail single-photon protocol at $\mathcal{R}_{\rm SR}\approx 0.11\sqrt{\eta}$.

% \begin{figure}[htbp]
% \centering
% \includegraphics[width=0.7\linewidth]{Fig2_v1.pdf}
% \caption{COW entanglement generation protocols with two different measurement strategies.}
% \label{fig:Fig2} 
% \end{figure} 

The second coherent-state-assisted entanglement generation protocol is named the coherent-one-way protocol (COW). Here, Alice generates a coherent state at amplitude $\alpha$, which interacts with her quantum memory first, as shown in Fig.~\ref{fig:Fig123}(d). The photonic state is transmitted over a lossy optical channel of transmissivity $\eta$ to Bob. After the interaction with the quantum memory at Bob's site, he carries out the state discrimination (to distinguish the phase of the coherent state). The first state discrimination scheme is the unambiguous state discrimination (USD)~\cite{duvsek2000unambiguous,mohseni2004optical,sidhu2023linear} between the two coherent states by using a local oscillator of matched amplitude $|\alpha\sqrt{\eta}\rangle$ and a 50:50 beamsplitter, followed by single-photon on-off detection of the output ports. This USD measurement projects the quantum memories into a Bell state depending on whether $D1$ or $D2$ clicks. The USD measurement COW protocol (labeled COW-USD with details in Appendix B1) yields an achievable distillable entanglement generation rate of
\begin{align}
    \mathcal{R}_{\rm COW}^{\rm USD}= \left(1-e^{-2\eta|\alpha|^2}\right)\left[1-h_2\left(\frac{1+e^{-2(1-\eta)|\alpha|^2}}{2}\right)\right]. 
\end{align}
The constituent terms in the hashing rate match closely with the CTW protocol and its hashing rate.
$\mathcal{R}_{\rm COW}^{\rm USD}$ approaches 1 ebit/ch at $\eta\rightarrow1$ and maintains $\mathcal{R}_{\rm COW}^{\rm USD}\approx 0.14\eta$ for $\eta\ll1$. Alternatively, Bob can also choose to perform the optimal state discrimination between two coherent states, which can be achieved by a Dolinar-type receiver, thus labeled as COW-DR (the third proposed protocol with details in Appendix B2). The Dolinar-type receiver applies a measurement-history-dependent adaptive displacement $D(\beta(t))$ to distinguish two coherent states and achieves an error rate approaching the Helstrom bound~\cite{dolinar1973optimum,cook2007optical,cui2022quantum}. This measurement always gives a {\em success} outcome. However, channel loss and the inherent probability of discrimination error induce bit-flip and phase-flip errors in the heralded state. For this scenario, the Hashing rate is given by
\begin{align}
    &\qquad \qquad \mathcal{R}_{\rm COW}^{\rm DR}=1 + \sum_{j,k=0,1} P_{j,k}\log_2(P_{j,k}),\nonumber\\
    &P_{j,k}=\frac{1 +(-1)^j e^{-2(1 - \eta)|\alpha|^2}}{2} \times \frac{1 +(-1)^k \sqrt{1 - e^{-4\eta|\alpha|^2}}}{2},
\end{align}
where subscripts $j,k=\{0,1\}$ indicate the occurrence of phase-flip (as a result of channel loss) and bit-flip errors (due to the non-zero discrimination error probability of the Dolinar receiver), respectively. 

% \begin{figure}[htbp]
% \centering
% \includegraphics[width=0.7\linewidth]{Fig3_v1.pdf}
% \caption{Evaluation of entanglement generation Hashing-bound rates of different protocols~\cite{dhara2024entangling,dhara2023entangling}. The advantageous area of coherent-state-assisted protocols is highlighted.}
% \label{fig:Fig3} 
% \end{figure} 

The rate-loss scaling curves of the three proposed protocols are evaluated by optimizing the initial coherent state amplitude $\alpha$ at each given end-to-end channel transmissivity $\eta$ and compared to the single-photon protocols in Fig.~\ref{fig:Fig123}(a). The proposed coherent-state-assisted protocols outperform the single-photon protocols in the low-loss regime ($\eta\lesssim0.5$ dB) while their performance degrades fast with loss. For $\eta\gtrsim0.5$ dB, they perform worse than the single-photon protocols. It is worth noting that the CTW and COW-USD protocols are unambiguous, which means they herald high-fidelity entanglement generation with sub-unity success probability when channel loss is present. Oppositely, the COW-DR protocol always returns a decision, but the channel loss induces more errors to the entangled state, which could potentially be corrected by measurement-based quantum error correction~\cite{shi2025measurement}.

\begin{figure}[htbp]
\centering
\includegraphics[width=\linewidth]{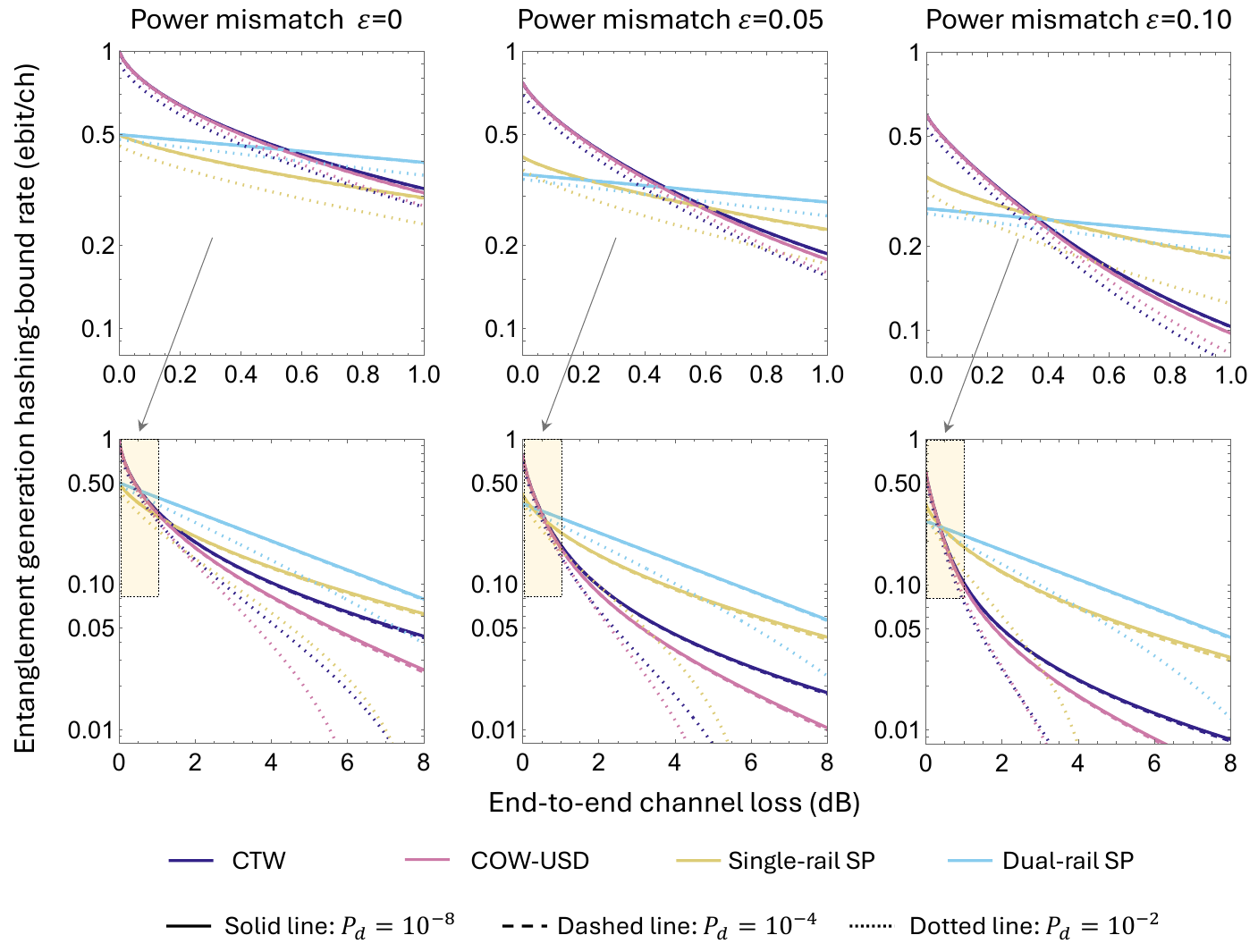}
\caption{Optimized hashing-bound entanglement generation rates of CTW, COW-USD, and single- and dual-rail single-photon (SP) protocols at different channel loss and imperfection settings. Columns correspond to power mismatch $\varepsilon=0$, $0.05$, and $0.10$. Solid, dashed, and dotted curves denote excess noise photon click probabilities $P_d=10^{-8}$, $10^{-4}$, and $10^{-2}$, respectively. The upper panels magnify the shaded low-loss regions of the lower panels, which use a logarithmic rate scale. The initial amplitude $\alpha$ is optimized at each coherent-state operating point.}
\label{fig:Fig4} 
\end{figure} 

\section{Impact of non-idealities}

In near-term experiments, other than losses, the primary sources of non-ideality that are likely to limit performance are excess noise in the channel (e.g., from the excess noise photon click probability of the detectors $P_d$), the sub-unity mode matching visibility (of the temporal-spectral mode), and imbalanced power arrived at the beamsplitter caused by imperfect calibration. We model excess photon noise by $P_d$ as it is independent of the initial coherent-state amplitude (Appendix~\ref{app:excess_noise}). Mode mismatch and imperfect calibration introduce remaining optical signal at the supposed dark port, whose mean photon number scales linearly with the input power. Therefore, we characterize this power-dependent noise using a single \textit{power mismatch} parameter $\varepsilon$, determined from the experimentally measured interference visibility. The conversion from conventional parameters calibrated in experiments to $\varepsilon$ is explained in Appendices~\ref{app:power_mismatch} and \ref{app:mode_matching}.

In the CTW protocol, excess noise leads to three possible errors: (1) $D1$ and $D2$ click at the same time, causing rejection of the event that decreases the overall success rate; (2) an extra photon at an already clicking detector changes the detected photon-number parity, causing a phase-flip error relative to the inferred Bell state; and (3) a dark count triggers either $D1$ or $D2$ when neither port contains an incoming signal photon, falsely accepting a separable vacuum-conditioned memory state. For COW-USD, the presence of excess noise reduces the overall entanglement generation rate, similar to the aforementioned cases (1) and (3) for the CTW protocol. Since COW-USD does not rely on the parity of PNR detection, the ebit/ch rate drop caused by the dark count is less significant compared to the CTW protocol. For both protocols, pulse power interference with sub-unity visibility decreases the success rate by reducing the chance of detecting a photon at the {\em bright} port (i.e., the port that is supposed to collect an output coherent state after a constructive interference) or triggering an ambiguous click at the {\em dark} port (i.e., the port that is supposed to obtain a vacuum output after a destructive interference). Detailed analyses of the protocol performance with non-ideal components are covered in Appendix~\ref{app:heralded_imperfect}.

We numerically evaluate the impact of these two non-idealities at selected values (labeled in the legend) in Fig.~\ref{fig:Fig4}. The inset plots in the top panel demonstrate that even with non-idealities, the ebit/ch rate advantage of the coherent-state-assisted protocols still holds in the low-loss regime.

\section{Toward Efficient Entanglement Generation among Multiple Quantum Memories}

Multipartite entanglement within a hub of closely linked quantum memories is a crucial resource for quantum error correction in a local quantum repeater infrastructure~\cite{dur2007entanglement,glaudell2016serialized,shi2025measurement}. Single-photon schemes based on unassisted linear-optical Bell measurements have an ideal success probability of at most $50\%$ per attempt. With suitable measurement bases and local corrections, heralded photon fusion failures in the ideal lossless setting can preserve previously established memory entanglement, allowing the addition of another memory to be retried~\cite{patil2026clifford}. The resulting multipartite-state preparation rate nevertheless depends on the number of repeated trials, error budget, and time cost on each elementary link. Boosted Bell-state measurements can increase the success probability, but require additional ancillary photons with well-mode-matched deterministic single photon sources~\cite{ewert2014efficient}.

\begin{figure}[htbp]
\centering
\includegraphics[width=0.6\linewidth]{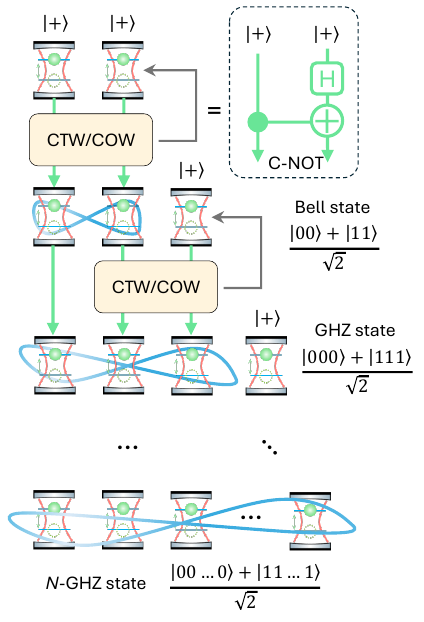}
\caption{A conceptual schematic for entangling multiple quantum memories into a GHZ state by iterating the CTW or COW protocol at the low-loss regime with feedback (with outcome-dependent local Pauli corrections).}
\label{fig:Fig5} 
\end{figure}

The proposed coherent-state-assisted protocols is equivalently a photon-mediated control-NOT (C-NOT) gate operating between two cavity-couple atomic or atomic-like systems, which increases the success probability of each two-qubit operation within the quantum memory hub in the low loss regime. As illustrated in Fig.~\ref{fig:Fig5}, these photon-mediated C-NOT gates enable the sequential preparation of an $N$-qubit GHZ state across $N$ quantum memories more deterministically than single-photon protocols, with at most twice fewer attempts on average for each successful addition. The increased success probability can therefore improve the overall multi-qubit entanglement preparation rate when operation times and the required final-state fidelity are taken into account, providing a new ingredient to future architectures for efficient and practical real-time resource-state generation in quantum repeaters.

\section{Conclusion and Discussion}

The three proposed coherent-state-assisted entanglement distribution protocols between two reflective-cavity-based quantum memories are near-term solutions that could boost both the ebit rate per channel use and success probability from 0.5 to 1.0, which surpasses the previous bound of using single-photon protocols in the low-loss regime. Based on this increased success probability, the protocols can improve the preparation rate of entangled states among more than two quantum memories, providing resource states to a local entanglement resource hub for quantum network applications~\cite{azuma2023quantum} and measurement-based quantum computing~\cite{briegel2009measurement,lanyon2013measurement,shi2025measurement}. Moreover, since the cavity-coupled
quantum memory supports both the single-photon protocols and the coherent-state-assisted protocols, such a monolithic platform may be enough to demonstrate the error-correction-empowered quantum repeater with constant overhead inside a quantum network~\cite{shi2025measurement}.

From the experimental perspective, since the coherent state can be easily generated and frequency-tuned by attenuating a laser after temporal-spectral mode engineering, the proposed protocols are relatively easier to implement in quantum memory systems in the very near term compared to the GKP state-assisted protocol ~\cite{dhara2024entangling} and QND protocol~\cite{winnel2022achieving}. Although there are still fabrication challenges toward the realization of a reflective-cavity-based quantum memory device and corresponding single-photon detectors with less than 0.5 dB net loss~\cite{reiserer2015cavity,sun2016quantum,sipahigil2016integrated,hacker2019deterministic,reddy2020superconducting,vasenin2024evolution}, these challenges will eventually and must be solved along the roadmap toward a scalable fault-tolerant quantum repeater and network architecture~\cite{awschalom2021development}. The results shown in this article may be a ladder to more novel protocols and seed practical demonstrations of generating highly entangled states in a local hub of quantum memories using phase-modulated coherent states. Further investigations along this direction with the underlying hardware parameter trade-offs at the qubit-photon interface~\cite{raymer2024duan} and error-rate thresholds toward fault tolerance~\cite{pant2019percolation} are open for future work.

\section*{Acknowledgments}
C.C. and S.G. acknowledge QuANET and PhENOM projects funded by DARPA. P.D. and S.G. acknowledge the Mega Qubit Router (MQR) project funded under federal support via a subcontract from the University of Arizona Applied Research Corporation (UA-ARC), for supporting this research. Additionally, all authors acknowledge the Engineering Research Center for Quantum Networks (CQN), awarded by the NSF and DoE under cooperative agreement number 1941583, for synergistic research support. 

\section*{Data availability}
The numerical data supporting the findings of this study are included in the figures. The analytical model and derivations are provided in the text and appendices.

\onecolumngrid    
\appendix
\section{Hashing Rate of the Coherent Two-Way (CTW) Protocol}

  \begin{figure}[htbp]
\centering
\includegraphics[width=0.35\linewidth]{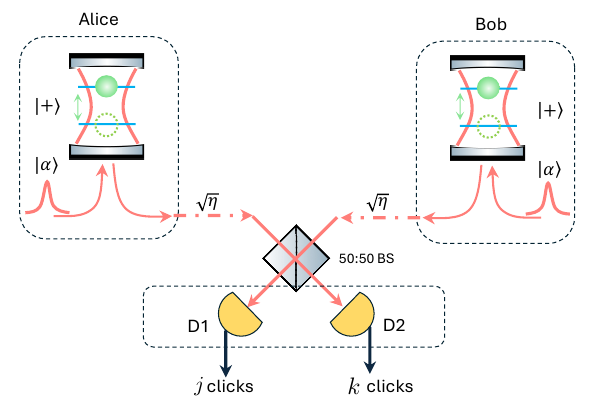}
\caption{Overview of the coherent two-way (CTW) protocol.}
\label{fig:supp_1} 
\end{figure} 

The initial coherent-state amplitude $\alpha$ and end-to-end intensity transmissivity $\eta$ follow the main-text definitions. Rates are measured in ebits per protocol attempt. We take $\alpha\in\mathbb{R}$ without loss of generality and use the coherent-state overlap
\begin{align}
    \braket{\alpha|\beta}=\exp\!\left(\alpha^*\beta-\tfrac{1}{2}\left(|\alpha|^2+|\beta|^2\right)\right).
\end{align}

The atom--cavity interaction imparts a memory-state-dependent phase rotation on a coherent pulse~\cite{duan2004scalable,reiserer2015cavity,raymer2024duan}. For a memory initialized in $(\ket{0}_M+\ket{1}_M)/\sqrt{2}$ and a coherent pulse $\ket{\alpha}_P$, the cavity--pulse interaction yields
\begin{align}
    \ket{\psi}_{MP}=\frac{1}{\sqrt{2}}\left( \ket{0}_M \ket{\alpha}_P +\ket{1}_M \ket{-\alpha}_P\right).
\end{align}
Transmission through a pure-loss channel of arm intensity transmissivity $\eta_r$ ($r=A,B$), with the lost light routed to an environment mode $E$, gives
\begin{align}
    \ket{\psi'}_{MPE}=\frac{1}{\sqrt{2}}\left( \ket{0}_M \ket{\alpha\sqrt{\eta_r}}_P \ket{\alpha\sqrt{1-\eta_r}}_E +\ket{1}_M \ket{-\alpha\sqrt{\eta_r}}_P \ket{-\alpha\sqrt{1-\eta_r}}_E\right).
\end{align}
Tracing out the environment yields
\begin{align}
    \rho_{MP} = \frac{1}{2}\sum_{i,j=0}^{1} \proj{i}{j}_M \otimes\proj{(-1)^i\alpha\sqrt{\eta_r}}{(-1)^j\alpha\sqrt{\eta_r}}_P  \times \Theta_{ij}(\alpha;\eta_r).
\end{align}
The environment overlap follows the main-text definition,
\begin{align}
 \Theta_{ij}(\alpha;\eta_r)
 =\braket{(-1)^j\alpha\sqrt{1-\eta_r}|(-1)^i\alpha\sqrt{1-\eta_r}}.
 \label{eq:environment_dephasing}
\end{align}
This overlap accounts for loss-induced memory dephasing. We retain the environment modes until calculating the conditional memory states.

In the CTW protocol, each party $r\in\{A,B\}$ holds one such state and the two pulses propagate through channels of transmissivity $\eta_A$ and $\eta_B$ to a central station. The joint state is
\begin{align}
    \ket{\psi'}^{(A)}_{MPE}\otimes\ket{\psi'}^{(B)}_{MPE}=\frac{1}{2}\sum_{i,j=0}^{1} \ket{i,j}_M \Ket{(-1)^i\alpha \sqrt{\eta_{A}},\,(-1)^j\alpha\sqrt{\eta_{B}}}_P \Ket{(-1)^i\alpha \sqrt{1-\eta_{A}},\,(-1)^j\alpha \sqrt{1-\eta_{B}}}_E,
\end{align}
where the first (second) entry of each two-mode ket refers to Alice's (Bob's) mode. Mixing the two pulse modes on a balanced beamsplitter, whose output ports are monitored by detectors $D_1$ and $D_2$, gives
\begin{align}
    \ket{\varphi} = \frac{1}{2}\sum_{i,j=0}^{1} \ket{i,j}_M \Ket{\beta^{+}_{ij},\,\beta^{-}_{ij}}_{D_1,D_2} \Ket{(-1)^i\alpha \sqrt{1-\eta_{A}},\,(-1)^j\alpha \sqrt{1-\eta_{B}}}_E,
    \qquad
    \beta^{\pm}_{ij}= \frac{(-1)^i \alpha\sqrt{\eta_{A}}\pm(-1)^j\alpha\sqrt{\eta_{B}}}{\sqrt{2}}.
\end{align}
We assume balanced arms, $\eta_A=\eta_B=\sqrt{\eta}$, so that the total link transmissivity is $\eta_A\eta_B=\eta$. Introducing the shorthand
\begin{align}
    \lambda\equiv\alpha\sqrt{1-\sqrt{\eta}}\quad\text{(environment amplitude)},\qquad
    \kappa\equiv\alpha\sqrt{2\sqrt{\eta}}\quad\text{(bright-port amplitude)},
\end{align}
the post-interference state reads
\begin{align}
    \begin{split}
        \ket{\varphi} = \frac{1}{2} \biggl( & \ket{0,0}_M \ket{\kappa,0}_{D_1,D_2} \ket{\lambda,\lambda}_E + \ket{1,1}_M \ket{-\kappa,0}_{D_1,D_2} \ket{-\lambda,-\lambda}_E \\
        &+ \ket{0,1}_M \ket{0,\kappa}_{D_1,D_2} \ket{\lambda,-\lambda}_E + \ket{1,0}_M \ket{0,-\kappa}_{D_1,D_2} \ket{-\lambda,\lambda}_E \biggr).
    \end{split}
    \label{eq:post_int_CTW}
\end{align}
Expanding the detected coherent states in the even and odd Schr\"odinger-cat basis gives
\begin{align}
    \ket{\mu_{e}(\kappa)}= \frac{\ket{\kappa} +\ket{-\kappa}}{\mathcal{N}_e},\qquad
    \ket{\mu_{o}(\kappa)}= \frac{\ket{\kappa} -\ket{-\kappa}}{\mathcal{N}_o},\qquad
    \mathcal{N}_{e(o)}=\sqrt{2\left(1\pm e^{-2|\kappa|^2}\right)},
\end{align}
so that $\ket{\pm \kappa} = \left(\mathcal{N}_e\ket{\mu_e(\kappa)} \pm \mathcal{N}_o\ket{\mu_o(\kappa)}\right)/2$. Grouping the terms of Eq.~\eqref{eq:post_int_CTW} by the cat state occupying the detected modes,
\begin{align}
    \begin{split}
        \ket{\varphi} = \frac{1}{4} \biggl( & \Bigl[\ket{0,0}_M \ket{\lambda,\lambda}_E + \ket{1,1}_M \ket{-\lambda,-\lambda}_E\Bigr]\mathcal{N}_e\ket{\mu_e(\kappa),0}_{D_1,D_2}
        \\ &+ \Bigl[\ket{0,0}_M \ket{\lambda,\lambda}_E - \ket{1,1}_M \ket{-\lambda,-\lambda}_E\Bigr]\mathcal{N}_o\ket{\mu_o(\kappa),0}_{D_1,D_2} \\
        &+ \Bigl[\ket{0,1}_M \ket{\lambda,-\lambda}_E + \ket{1,0}_M \ket{-\lambda,\lambda}_E\Bigr]\mathcal{N}_e\ket{0,\mu_e(\kappa)}_{D_1,D_2} \\
        &+ \Bigl[\ket{0,1}_M \ket{\lambda,-\lambda}_E - \ket{1,0}_M \ket{-\lambda,\lambda}_E\Bigr]\mathcal{N}_o\ket{0,\mu_o(\kappa)}_{D_1,D_2}\biggr).
    \end{split}
\end{align}
The detectors perform photon-number-resolving measurements with POVM elements $\hat{\Pi}_{j,k} = \outprod{j}_{D_1}\otimes \outprod{k}_{D_2}$, $j,k \in \mathbb{Z}_{\geq 0}$. Since $\ket{\mu_e(\kappa)}$ ($\ket{\mu_o(\kappa)}$) has support only on even (odd) photon numbers, the outcomes fall into five classes. Let $\bar n_{D_b}\equiv|\kappa|^2 = 2\sqrt{\eta}|\alpha|^2$ denote the mean photon number in the bright port, with $b=1$ for even memory parity and $b=2$ for odd memory parity; the probability of registering $j>0$ photons at $D_1$ and none at $D_2$ is $e^{-\bar n_{D_b}}\bar n_{D_b}^{j}/(2\,j!)$, and analogously with the roles of $D_1$ and $D_2$ exchanged. Summing over the relevant parities gives:
\begin{enumerate}
    \item $k=0$ and $j>0$ even: occurs with probability
    \begin{align}
        P_{e}(\bar n_{D_b}) \equiv \frac{\left(1-e^{-\bar n_{D_b}}\right)^2}{4}
    \end{align}
    and heralds
    \begin{align}
        \ket{\varphi_1} =  \left(\ket{0,0}_M \ket{\lambda,\lambda}_E  + \ket{1,1}_M \ket{-\lambda,-\lambda}_E \right)/\sqrt{2}.
    \end{align}
    \item $k=0$ and $j$ odd: occurs with probability
    \begin{align}
        P_{o}(\bar n_{D_b}) \equiv \frac{1-e^{-2\bar n_{D_b}}}{4}
    \end{align}
    and heralds
    \begin{align}
        \ket{\varphi_2} =  \left(\ket{0,0}_M \ket{\lambda,\lambda}_E  - \ket{1,1}_M \ket{-\lambda,-\lambda}_E \right)/\sqrt{2}.
    \end{align}
    \item $j=0$ and $k>0$ even: occurs with probability $P_e(\bar n_{D_b})$ and heralds
    \begin{align}
        \ket{\varphi_3} =  \left(\ket{0,1}_M \ket{\lambda,-\lambda}_E  + \ket{1,0}_M \ket{-\lambda,\lambda}_E \right)/\sqrt{2}.
    \end{align}
    \item $j=0$ and $k$ odd: occurs with probability $P_o(\bar n_{D_b})$ and heralds
    \begin{align}
        \ket{\varphi_4} =  \left(\ket{0,1}_M \ket{\lambda,-\lambda}_E  - \ket{1,0}_M \ket{-\lambda,\lambda}_E \right)/\sqrt{2}.
    \end{align}
    \item $j=k=0$: occurs with probability $P_f(\bar n_{D_b})\equiv e^{-\bar n_{D_b}}$ and heralds
    \begin{align}
        \ket{\varphi_5} =  \Bigl( \ket{0,0}_M \ket{\lambda,\lambda}_E  + \ket{1,1}_M \ket{-\lambda,-\lambda}_E +\ket{0,1}_M \ket{\lambda,-\lambda}_E  + \ket{1,0}_M \ket{-\lambda,\lambda}_E \Bigr)/2 .
    \end{align}
\end{enumerate}
Separating the vacuum outcome from the nonzero even-parity outcomes gives $2P_e+2P_o+P_f=1$. The state $\ket{\varphi_5}$ factorizes as $\bigl(\ket{0}_{M_A}\ket{\lambda}_{E_A}+\ket{1}_{M_A}\ket{-\lambda}_{E_A}\bigr)\otimes\bigl(\ket{0}_{M_B}\ket{\lambda}_{E_B}+\ket{1}_{M_B}\ket{-\lambda}_{E_B}\bigr)/2$ and is therefore separable across the memories. Consequently, the total probability of success is
\begin{align}
    P_{\rm CTW}(\alpha,\eta) = 2P_e+2P_o = 1-e^{-\bar n_{D_b}} = 1-\exp\!\left(-2\sqrt{\eta}|\alpha|^2\right).
\end{align}
    
The normalized conditional memory states are $\rho^{(i)}=\Tr_E(\outprod{\varphi_i})$, $i=1,\ldots,4$. Tracing out the environment from $\ket{\varphi_1}$ gives
\begin{align}
    \begin{split}
        \Tr_E\left(\outprod{\varphi_1}\right) &=  \frac{1}{2} \Bigl( \outprod{0,0}_M + \outprod{1,1}_M + \Theta_{\rm CTW}\left( \proj{0,0}{1,1}  +\proj{1,1}{0,0} \right) \Bigr)\\
        &= \frac{1+\Theta_{\rm CTW}}{2} \outprod{\Phi^+}_M +\frac{1-\Theta_{\rm CTW}}{2}\outprod{\Phi^-}_M \equiv \rho^{(1)},
    \end{split}
    \label{eq:CTW_rho_one}
\end{align}
The CTW loss-induced coherence factor is the product of the two arm environment overlaps,
\begin{align}
    \Theta_{\rm CTW} = \braket{-\lambda,-\lambda|\lambda,\lambda} = \Theta_{01}(\alpha;\sqrt\eta)^2 = e^{-4|\lambda|^2} = \exp\!\left(-4\left(1-\sqrt{\eta}\right)|\alpha|^2\right).
\end{align}
The remaining heralded states are
\begin{align}
    \begin{split}
        \Tr_E\left(\outprod{\varphi_2}\right) &= \frac{1+\Theta_{\rm CTW}}{2} \outprod{\Phi^-}_M +\frac{1-\Theta_{\rm CTW}}{2}\outprod{\Phi^+}_M \equiv \rho^{(2)},\\
        \Tr_E\left(\outprod{\varphi_3}\right) &= \frac{1+\Theta_{\rm CTW}}{2} \outprod{\Psi^+}_M +\frac{1-\Theta_{\rm CTW}}{2}\outprod{\Psi^-}_M \equiv \rho^{(3)},\\
        \Tr_E\left(\outprod{\varphi_4}\right) &= \frac{1+\Theta_{\rm CTW}}{2} \outprod{\Psi^-}_M +\frac{1-\Theta_{\rm CTW}}{2}\outprod{\Psi^+}_M \equiv \rho^{(4)}.
    \end{split}
    \label{eq:CTW_final_states}
\end{align}
Each of these Bell-diagonal states has the same hashing bound,
\begin{align}
    I(\rho^{(i)}) = 1-h_2\!\left(\frac{1+\Theta_{\rm CTW}}{2}\right) \equiv I_{\rm CTW} (\alpha,\eta),\qquad i \in\{1,2,3,4\},
\end{align}
where $h_2(p)=-p \log_2 p - (1-p) \log_2 (1-p)$ is the binary entropy function. The hashing rate of the CTW protocol is therefore
\begin{align}
    \begin{split}
        \mathcal{R}_{\mathrm{CTW}}(\alpha,\eta) &= P_{\rm CTW}(\alpha,\eta)\times I_{\rm CTW}(\alpha,\eta) \\
        &= \left(1-e^{-2\sqrt{\eta}|\alpha|^2}\right)\biggl[1-h_2\!\left(\frac{1+e^{-4(1-\sqrt{\eta})|\alpha|^2}}{2}\right)\biggr].
    \end{split}
\end{align}

\section{Hashing Rates of the Coherent One-Way (COW) Protocols}

\begin{figure}[htbp]
\centering
\includegraphics[width=0.35\linewidth]{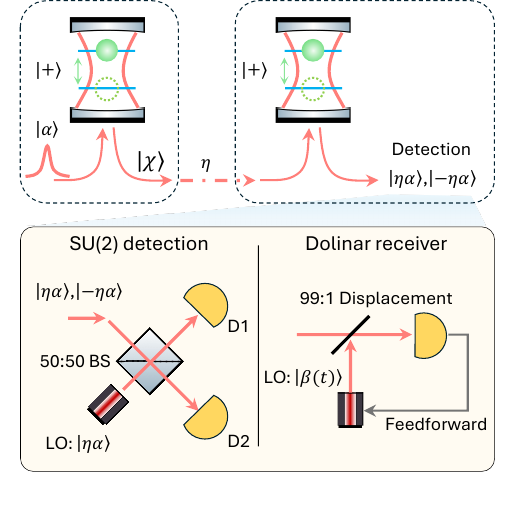}
\caption{Overview of the coherent one-way (COW) protocols.}
\label{fig:supp_2} 
\end{figure} 

In the coherent one-way (COW) protocol, a single traveling coherent pulse interacts sequentially with the two quantum memories. Without loss of generality, Alice's memory interacts first, producing $\ket{\psi}_{MP}=\left( \ket{0}_M \ket{\alpha}_P +\ket{1}_M \ket{-\alpha}_P\right)/\sqrt{2}$ as above. Transmission over a pure-loss channel of transmissivity $\eta$ gives
\begin{align}
    \ket{\psi'}^{(A)}_{MPE} = \frac{1}{\sqrt{2}}\left( \ket{0}_M \ket{\alpha\sqrt{\eta}}_P \ket{\alpha\sqrt{1-\eta}}_E +\ket{1}_M \ket{-\alpha\sqrt{\eta}}_P \ket{-\alpha\sqrt{1-\eta}}_E\right),
\end{align}
and the subsequent interaction with Bob's memory yields, with $\lambda'\equiv\alpha\sqrt{1-\eta}$,
\begin{align}
    \begin{split}
        \ket{\psi''}^{(AB)}_{MPE} = \frac{1}{2}\biggl( & \ket{0,0}_M \ket{\alpha \sqrt{\eta}}_P \ket{\lambda'}_E 
        + \ket{0,1}_M \ket{-\alpha \sqrt{\eta}}_P \ket{\lambda'}_E \\
        & + \ket{1,0}_M \ket{-\alpha \sqrt{\eta}}_P \ket{-\lambda'}_E
        + \ket{1,1}_M \ket{\alpha \sqrt{\eta}}_P \ket{-\lambda'}_E \biggr).
    \end{split}
    \label{eq:COW_pre_int_SM}
\end{align}
The pulse phase now encodes the two-memory parity. We consider two readout schemes: unambiguous state discrimination (USD) and the Dolinar receiver.

\subsection{Unambiguous State Discrimination}
In the USD approach, the pulse is interfered on a balanced beamsplitter with a local-oscillator (LO) pulse $\ket{\alpha\sqrt{\eta}}_{\mathrm{LO}}$. Writing $\kappa'\equiv\alpha\sqrt{2\eta}$ for the bright-port amplitude, Eq.~\eqref{eq:COW_pre_int_SM} becomes
\begin{align}
    \begin{split}
        \ket{\phi}^{(AB)} = \frac{1}{2}\biggl( & \Bigl(\ket{0,0}_M  \ket{\lambda'}_E  + \ket{1,1}_{M} \ket{-\lambda'}_E \Bigr) \ket{\kappa',0}_{D_1,D_2} 
        + \Bigl(\ket{0,1}_M \ket{\lambda'}_E  + \ket{1,0}_{M} \ket{-\lambda'}_E \Bigr) \ket{0,-\kappa'}_{D_1,D_2} \biggr).
    \end{split}
    \label{eq:post_int_COW_SM}
\end{align}
On--off photodetection of the two output ports yields one of three outcomes:
\begin{enumerate}
    \item $D_1$ clicks and $D_2$ does not. This occurs with probability $P_{s}\equiv\left(1-e^{-2\eta|\alpha|^2}\right)/2$ and heralds
    \begin{align}
        \ket{\chi_1} = \frac{\ket{0,0}_M  \ket{\lambda'}_E  + \ket{1,1}_{M} \ket{-\lambda'}_E }{\sqrt{2}}.
    \end{align}
    \item $D_2$ clicks and $D_1$ does not. This occurs with probability $P_s$ and heralds
    \begin{align}
        \ket{\chi_2} = \frac{\ket{0,1}_M  \ket{\lambda'}_E  + \ket{1,0}_{M} \ket{-\lambda'}_E }{\sqrt{2}}.
    \end{align}
    \item Neither detector clicks. This occurs with probability $P_{f}\equiv e^{-2\eta|\alpha|^2}=1-2P_s$ and heralds the separable state
    \begin{align}
        \ket{\chi_3} = \frac{ \ket{0,0}_M  \ket{\lambda'}_E  + \ket{1,1}_{M} \ket{-\lambda'}_E + \ket{0,1}_M  \ket{\lambda'}_E  + \ket{1,0}_{M} \ket{-\lambda'}_E }{2}.
    \end{align}
\end{enumerate}
The two single-click outcomes are accepted, so the probability of success is $P_{\rm COW, USD}(\alpha,\eta) = 2P_s = 1- e^{-2\eta|\alpha|^2}$. Tracing out the environment,
\begin{align}
    \begin{split}
        \Tr_E\left(\outprod{\chi_1}\right) &= \frac{1+\Theta_{\rm COW}}{2} \outprod{\Phi^+}_M + \frac{1-\Theta_{\rm COW}}{2} \outprod{\Phi^-}_M \equiv \sigma^{(1)},\\
        \Tr_E\left(\outprod{\chi_2}\right) &= \frac{1+\Theta_{\rm COW}}{2} \outprod{\Psi^+}_M + \frac{1-\Theta_{\rm COW}}{2} \outprod{\Psi^-}_M  \equiv \sigma^{(2)},
    \end{split}
    \label{eq:su2_outcome}
\end{align}
where $\Theta_{\rm COW}=\Theta_{01}(\alpha;\eta)=e^{-2|\lambda'|^2}=\exp\!\left(-2(1-\eta)|\alpha|^2\right)$. The single environment mode sets the COW dephasing factor. The states $\sigma^{(1)}$ and $\sigma^{(2)}$ correspond to even and odd memory parity, respectively, and share the hashing bound
\begin{align}
    I(\sigma^{(j)}) = 1-h_2\!\left(\frac{1+\Theta_{\rm COW}}{2}\right) \equiv I_{\rm COW, USD}(\alpha,\eta),\qquad  j \in\{1,2\},
\end{align}
and the hashing rate of the COW protocol with USD detection is
\begin{align}
    \begin{split}
        \mathcal{R}^{\mathrm{USD}}_{\mathrm{COW}}(\alpha,\eta) &= P_{\rm COW, USD}(\alpha,\eta)\times I_{\rm COW, USD}(\alpha,\eta)\\
        &= \left(1-e^{-2\eta|\alpha|^2}\right)\biggl[1-h_2\!\left(\frac{1+e^{-2(1-\eta)|\alpha|^2}}{2}\right)\biggr].
    \end{split}
\end{align}

\subsection{Dolinar Receiver}
The Dolinar receiver is the optimal receiver for discriminating binary phase-shift-keyed coherent states in classical optical communication. It applies a measurement-history-dependent adaptive displacement $D(\beta(t))$ and achieves the Helstrom bound: for the equiprobable states $\ket{\pm\nu}$, the minimum error probability is $p_{\rm err} = \bigl(1-\sqrt{1-e^{-4|\nu|^2}}\bigr)/2$. 

From Eq.~\eqref{eq:COW_pre_int_SM}, the states $\ket{\alpha\sqrt{\eta}}_P$ and $\ket{-\alpha\sqrt{\eta}}_P$ are equally likely, so $\nu=\alpha\sqrt{\eta}$ and
\begin{align}
    p_{\rm err}(\alpha,\eta) = \frac{1-\sqrt{1-e^{-4\eta|\alpha|^2}}}{2}.
\end{align}
The Dolinar receiver returns a decision in every round. To obtain an achievable Bell-diagonal hashing rate, we apply bilateral Pauli twirling to its conditional memory state: the same randomly chosen Pauli operator is applied to both memories and the random label is discarded. This removes coherence between Bell states and gives, in terms of the normalized USD states $\sigma^{(1)}$ and $\sigma^{(2)}$ defined in Eq.~\eqref{eq:su2_outcome},
\begin{subequations}
    \begin{align}
        &\text{Decision ``}\ket{\alpha\sqrt{\eta}}\text{''} : \varsigma^{(1)} = (1-p_{\rm err})\,\sigma^{(1)} + p_{\rm err}\,\sigma^{(2)},\\
        &\text{Decision ``}\ket{-\alpha\sqrt{\eta}}\text{''} : \varsigma^{(2)} = (1-p_{\rm err})\,\sigma^{(2)} + p_{\rm err}\,\sigma^{(1)}.
    \end{align}
\end{subequations}
The two Bell-diagonal states share an eigenvalue spectrum that factors into loss-induced phase-flip and receiver-induced bit-flip probabilities,
\begin{align}
    P_{j,k} =\frac{1 +(-1)^j e^{-2(1 - \eta)|\alpha|^2}}{2} \times \frac{1 +(-1)^k \sqrt{1 - e^{-4\eta|\alpha|^2}}}{2}, \qquad j,k\in\{0,1\}.
\end{align}
Since every round yields a decision, the achievable rate is the hashing bound of the twirled state,
\begin{align}
    \mathcal{R}^{\mathrm{DR}}_{\mathrm{COW}}(\alpha,\eta) = \max\!\left\{0,\,1 + \sum_{j,k=0}^{1} P_{j,k} \log_2 P_{j,k}\right\}.
\end{align}

\section{Protocol Performance with Non-Idealities}
Input power imbalance, imperfect mode matching, and phase drift reduce interference contrast and can be converted to a single power-mismatch parameter $\varepsilon$. Background photons and detector dark counts contribute the power-independent excess noise photon click probability $P_d$. For each memory-parity sector, $D_b$ and $D_d$ denote the nominally bright and dark ports, with $(b,d)=(1,2)$ for even parity and $(2,1)$ for odd parity. In terms of their optical mean photon numbers, we define
\begin{align}
    \varepsilon=\frac{\bar n_{D_d}}{\bar n_{D_b}+\bar n_{D_d}}.
    \label{eq:common_output_mismatch}
\end{align}

\subsection{Input Power Mismatch}
\label{app:power_mismatch}
Consider two pulses with arriving field amplitudes $\alpha_1,\alpha_2\geq0$ in the same optical mode, with their relative phase stabilized. Interference on a balanced beamsplitter gives
\begin{align}
    \ket{\alpha_1}_A\ket{\alpha_2}_B\longmapsto
    \Ket{\frac{\alpha_1+\alpha_2}{\sqrt2}}_{D_1}
    \Ket{\frac{\alpha_1-\alpha_2}{\sqrt2}}_{D_2}.
\end{align}
For the even-memory-parity sector, the mean photon numbers are $\bar n_{D_b}=(\alpha_1+\alpha_2)^2/2$ and $\bar n_{D_d}=(\alpha_1-\alpha_2)^2/2$; the ports exchange roles for odd parity. Thus input power imbalance can be converted to
\begin{align}
    \varepsilon=\frac{(\alpha_1-\alpha_2)^2}{2(\alpha_1^2+\alpha_2^2)}.
\end{align}
Unequal arm loss enters through the arriving amplitudes. For equal initial pulse powers and arm intensity transmissivities $\eta_A,\eta_B$, with $\eta=\eta_A\eta_B$, the conversion is $\varepsilon=1/2-\sqrt{\eta_A\eta_B}/(\eta_A+\eta_B)$.

\subsection{Imperfect Mode Matching}
\label{app:mode_matching}
Imperfect mode overlap can be converted to the same power-mismatch parameter $\varepsilon$. We first consider spectral mismatch with a stabilized relative phase; the treatment also applies to unresolved polarization or transverse spatial modes.

Consider two pulses in coherent states $\ket{\alpha_1}_A$ and $\ket{\alpha_2}_B$ with spectral mode functions $f_A(\omega)$ and $f_B(\omega)$, normalized in $L^2(\mathbb{R})$. Defining the mode creation operators $A_f^\dagger=\int_{-\infty}^{\infty} d\omega\, f(\omega)\, a^\dagger(\omega)$, the pulses are
\begin{align}
    \ket{\alpha_1}_A =  \exp\!\left(\alpha_1 A_{f_A}^\dagger - \alpha_1^* A_{f_A}\right)\ket{0},\qquad
    \ket{\alpha_2}_B =  \exp\!\left(\alpha_2 A_{f_B}^\dagger - \alpha_2^* A_{f_B}\right)\ket{0}.
\end{align}
Perfectly matched modes satisfy $\left|\int d\omega\, f_B^*(\omega) f_A(\omega)\right| = 1$. In general we write $\int d\omega\, f_B^*(\omega) f_A(\omega) = \sqrt{\mathcal{V}}\,e^{i\theta}$ with squared mode overlap $\mathcal{V}\in[0,1]$; absorbing the phase $\theta$ into the definition of $f_B$, we take the overlap to be real. For $\mathcal V<1$, a Gram--Schmidt orthogonalization yields the orthonormal functions
\begin{align}
    g_1(\omega) = f_A(\omega), \qquad g_2(\omega) = \frac{f_B(\omega) - \sqrt{\mathcal{V}}\, f_A(\omega)}{\sqrt{1-\mathcal{V}}},
\end{align}
in terms of which the (spatially separated) input pulses are
\begin{align}
    \ket{\alpha_1}_A = \ket{\alpha_1}_{A,g_1}\otimes \ket{0}_{A,g_2}, \qquad
    \ket{\alpha_2}_B = \ket{\alpha_2\sqrt{\mathcal{V}}}_{B,g_1}\otimes \ket{\alpha_2\sqrt{1-\mathcal{V}}}_{B,g_2},
\end{align}
where the labels $g_1$ and $g_2$ denote the matched and unmatched portions of each pulse. Interference on a balanced beamsplitter yields, in the output modes $D_1$ and $D_2$,
\begin{align}
    \begin{split}
        & \Ket{\frac{\alpha_1+\alpha_2\sqrt{\mathcal{V}}}{\sqrt{2}}}_{D_1,g_1} \otimes \Ket {\frac{\alpha_2\sqrt{1-\mathcal{V}}}{\sqrt{2}}}_{D_1,g_2}, \\
        & \Ket{\frac{\alpha_1-\alpha_2\sqrt{\mathcal{V}}}{\sqrt{2}}}_{D_2,g_1} \otimes \Ket {\frac{-\alpha_2\sqrt{1-\mathcal{V}}}{\sqrt{2}}}_{D_2,g_2}.
    \end{split}
\end{align}

The mean photon number at each output port is the sum of the squared magnitudes of its two spectral-mode amplitudes. At equal arriving powers, the fraction in the nominally dark port is
\begin{align}
    \varepsilon=\frac{1-\sqrt{\mathcal V}}{2}.
\end{align}

To include phase drift, let $\phi_t$ be the time-dependent relative optical phase after calibration of its offset, and let $f_A(t),f_B(t)$ be the normalized temporal pulse envelopes, with $\int dt\,|f_A(t)|^2=\int dt\,|f_B(t)|^2=1$. The final power mismatch is
\begin{align}
    \varepsilon=\frac12-\frac{\alpha_1\alpha_2}{\alpha_1^2+\alpha_2^2}
    \operatorname{Re}\!\int dt\,f_A^*(t)f_B(t)e^{i\phi_t}.
\end{align}
This also includes simultaneous input power imbalance and mode mismatch. For equal arriving powers and identical envelopes $f(t)$, it becomes $\varepsilon=[1-\int dt\,|f(t)|^2\cos\phi_t]/2$. For phase drift or a random walk across attempts, the unnormalized conditional states below are evaluated for each trajectory and averaged before normalization and evaluation of the hashing rate.

\subsection{Amplitude-Independent Excess Noise}
\label{app:excess_noise}
Excess noise from background photons or dark counts changes the accepted detector patterns and mixes the corresponding conditional memory states. Let $P_d$ be the power-independent excess noise photon click probability at each detector during a heralding window. We model independent Bernoulli noise at the two detectors, with one added count per noise event. This approximates Poisson dark counts to first order in $P_d\ll1$.

For photon-number-resolving detection, an added count flips the registered photon-number parity. For on--off detection, an added count at an already-clicking port preserves the click pattern. An added count at the other port causes rejection, whereas a count when both optical outputs are vacuum can produce a false herald.

\subsection{Heralded Entangled State under Imperfections}
\label{app:heralded_imperfect}
Let $j,k\in\mathbb Z_{\geq0}$ be the registered photon counts at $D_1,D_2$, including counts from excess noise. After bilateral Pauli twirling, as in the Dolinar analysis, each heralded state is a mixture of the ideal memory states derived above. We write the von Neumann entropy as $H(\rho)=-\Tr(\rho\log_2\rho)$.

\paragraph{CTW protocol.}
For symmetric arms, the optical mean photon numbers and loss-induced coherence factor are
\begin{align}
    \bar n_{D_b}&=2\sqrt\eta|\alpha|^2(1-\varepsilon),\qquad
    \bar n_{D_d}=2\sqrt\eta|\alpha|^2\varepsilon,\\
    \Theta_{\rm CTW}&=\Theta_{01}(\alpha;\sqrt\eta)^2=e^{-4(1-\sqrt\eta)|\alpha|^2}.
\end{align}
We use the functions $P_e$ and $P_o$ and the normalized memory states $\rho^{(i)}=\Tr_E(\outprod{\varphi_i})$, $i=1,\ldots,4$, defined in Eqs.~\eqref{eq:CTW_rho_one} and \eqref{eq:CTW_final_states}. Conditioning on vacuum in both detected optical modes gives the memory--environment state $\ket{\varphi_5}$ from Appendix A. Tracing out the environment and applying bilateral Pauli twirling gives the normalized memory state $\rho^{(5)}=(I_4+\Theta_{\rm CTW} X\otimes X)/4$, where $I_4$ is the two-memory identity and $X$ is the Pauli operator. For each accepted class, $R^{(i)}(\varepsilon,P_d)$ denotes its unnormalized memory state; its trace is the probability of that class.

\begin{enumerate}
\item $k=0$ and $j>0$ even:
\begin{align}
 R^{(1)}(\varepsilon,P_d)
 &= (1-P_d)^2\bigl[P_e(\bar n_{D_b})e^{-\bar n_{D_d}}\rho^{(1)}
                  +P_e(\bar n_{D_d})e^{-\bar n_{D_b}}\rho^{(3)}\bigr]\nonumber\\
 &\quad+P_d(1-P_d)\bigl[P_o(\bar n_{D_b})e^{-\bar n_{D_d}}\rho^{(2)}
                  +P_o(\bar n_{D_d})e^{-\bar n_{D_b}}\rho^{(4)}\bigr].
\end{align}
\item $k=0$ and $j$ odd:
\begin{align}
 R^{(2)}(\varepsilon,P_d)
 &= (1-P_d)^2\bigl[P_o(\bar n_{D_b})e^{-\bar n_{D_d}}\rho^{(2)}
                  +P_o(\bar n_{D_d})e^{-\bar n_{D_b}}\rho^{(4)}\bigr]\nonumber\\
 &\quad+P_d(1-P_d)\bigl[P_e(\bar n_{D_b})e^{-\bar n_{D_d}}\rho^{(1)}
                  +P_e(\bar n_{D_d})e^{-\bar n_{D_b}}\rho^{(3)}
                  +e^{-(\bar n_{D_b}+\bar n_{D_d})}\rho^{(5)}\bigr].
\end{align}
\item $j=0$ and $k>0$ even:
\begin{align}
 R^{(3)}(\varepsilon,P_d)
 &= (1-P_d)^2\bigl[P_e(\bar n_{D_b})e^{-\bar n_{D_d}}\rho^{(3)}
                  +P_e(\bar n_{D_d})e^{-\bar n_{D_b}}\rho^{(1)}\bigr]\nonumber\\
 &\quad+P_d(1-P_d)\bigl[P_o(\bar n_{D_b})e^{-\bar n_{D_d}}\rho^{(4)}
                  +P_o(\bar n_{D_d})e^{-\bar n_{D_b}}\rho^{(2)}\bigr].
\end{align}
\item $j=0$ and $k$ odd:
\begin{align}
 R^{(4)}(\varepsilon,P_d)
 &= (1-P_d)^2\bigl[P_o(\bar n_{D_b})e^{-\bar n_{D_d}}\rho^{(4)}
                  +P_o(\bar n_{D_d})e^{-\bar n_{D_b}}\rho^{(2)}\bigr]\nonumber\\
 &\quad+P_d(1-P_d)\bigl[P_e(\bar n_{D_b})e^{-\bar n_{D_d}}\rho^{(3)}
                  +P_e(\bar n_{D_d})e^{-\bar n_{D_b}}\rho^{(1)}
                  +e^{-(\bar n_{D_b}+\bar n_{D_d})}\rho^{(5)}\bigr].
\end{align}
\item $j=k=0$: occurs with probability $(1-P_d)^2e^{-(\bar n_{D_b}+\bar n_{D_d})}$ and yields the separable state $\rho^{(5)}$; this class is discarded.
\item $j>0$ and $k>0$: occurs with probability $[1-(1-P_d)e^{-\bar n_{D_b}}][1-(1-P_d)e^{-\bar n_{D_d}}]$; this class is discarded.
\end{enumerate}
The $e^{-\bar n_{D_b}}$ terms describe dark-port counts accepted with the wrong memory parity. The $P_d(1-P_d)$ terms describe noise-induced parity flips and false heralds when both optical outputs are vacuum. For $i=1,\ldots,4$, the class probabilities, normalized states, and hashing rate are
\begin{align}
 p_i(\varepsilon,P_d)&=\Tr R^{(i)}(\varepsilon,P_d),\qquad
 \widetilde\rho^{(i)}(\varepsilon,P_d)=\frac{R^{(i)}(\varepsilon,P_d)}{p_i(\varepsilon,P_d)},\\
 \mathcal R_{\rm CTW}(\alpha,\eta;\varepsilon,P_d)
 &=\sum_{i:p_i>0}p_i(\varepsilon,P_d)
       \max\{0,1-H[\widetilde\rho^{(i)}(\varepsilon,P_d)]\}.
\end{align}

\paragraph{COW--USD protocol.}
The corresponding optical mean photon numbers and loss-induced coherence factor are
\begin{align}
 \bar n_{D_b}&=2\eta|\alpha|^2(1-\varepsilon),\qquad
 \bar n_{D_d}=2\eta|\alpha|^2\varepsilon,\\
 \Theta_{\rm COW}&=\Theta_{01}(\alpha;\eta)=e^{-2(1-\eta)|\alpha|^2}.
\end{align}
We use the normalized USD states $\sigma^{(1)},\sigma^{(2)}$ from Eq.~\eqref{eq:su2_outcome}. Conditioning on vacuum in both detected optical modes gives the memory--environment state $\ket{\chi_3}$ from Appendix B1. Tracing out the environment and applying bilateral Pauli twirling gives the normalized memory state $\sigma^{(3)}=(I_4+\Theta_{\rm COW}X\otimes X)/4$. With on--off detection, the registered counts determine the following click classes:
\begin{enumerate}
\item $j>0$, $k=0$ ($D_1$ clicks and $D_2$ does not):
\begin{align}
 S^{(1)}(\varepsilon,P_d)
 &=\frac{1-P_d}{2}\bigl[(1-e^{-\bar n_{D_b}})e^{-\bar n_{D_d}}\sigma^{(1)}
      +(1-e^{-\bar n_{D_d}})e^{-\bar n_{D_b}}\sigma^{(2)}\bigr]\nonumber\\
 &\quad+P_d(1-P_d)e^{-(\bar n_{D_b}+\bar n_{D_d})}\sigma^{(3)}.
\end{align}
\item $j=0$, $k>0$ ($D_2$ clicks and $D_1$ does not):
\begin{align}
 S^{(2)}(\varepsilon,P_d)
 &=\frac{1-P_d}{2}\bigl[(1-e^{-\bar n_{D_b}})e^{-\bar n_{D_d}}\sigma^{(2)}
      +(1-e^{-\bar n_{D_d}})e^{-\bar n_{D_b}}\sigma^{(1)}\bigr]\nonumber\\
 &\quad+P_d(1-P_d)e^{-(\bar n_{D_b}+\bar n_{D_d})}\sigma^{(3)}.
\end{align}
\item $j=k=0$: occurs with probability $(1-P_d)^2e^{-(\bar n_{D_b}+\bar n_{D_d})}$ and yields the separable state $\sigma^{(3)}$; this class is discarded.
\item $j>0$ and $k>0$: occurs with probability $[1-(1-P_d)e^{-\bar n_{D_b}}][1-(1-P_d)e^{-\bar n_{D_d}}]$; this class is discarded.
\end{enumerate}
An added count at an already-clicking port preserves the click pattern. The factor $1-P_d$ in the first terms accounts for the absence of an added count at the other port. The final terms describe false single clicks when both optical outputs are vacuum. For $\ell=1,2$, the class probabilities, normalized states, and rate are
\begin{align}
 q_\ell(\varepsilon,P_d)&=\Tr S^{(\ell)}(\varepsilon,P_d),\qquad
 \widetilde\sigma^{(\ell)}(\varepsilon,P_d)=\frac{S^{(\ell)}(\varepsilon,P_d)}{q_\ell(\varepsilon,P_d)},\\
 \mathcal R_{\rm COW}^{\rm USD}(\alpha,\eta;\varepsilon,P_d)
 &=\sum_{\ell:q_\ell>0}q_\ell(\varepsilon,P_d)
       \max\{0,1-H[\widetilde\sigma^{(\ell)}(\varepsilon,P_d)]\}.
\end{align}

\twocolumngrid
\bibliography{updated_ref}

@article{azuma2023quantum,
  title={Quantum repeaters: From quantum networks to the quantum internet},
  author={Azuma, Koji and Economou, Sophia E and Elkouss, David and Hilaire, Paul and Jiang, Liang and Lo, Hoi-Kwong and Tzitrin, Ilan},
  journal={Reviews of Modern Physics},
  volume={95},
  number={4},
  pages={045006},
  year={2023},
  publisher={APS},
  doi={10.1103/RevModPhys.95.045006},
  url={https://doi.org/10.1103/RevModPhys.95.045006}
}

@article{briegel1998quantum,
  title={Quantum repeaters: the role of imperfect local operations in quantum communication},
  author={Briegel, H-J and D{\"u}r, Wolfgang and Cirac, Juan I and Zoller, Peter},
  journal={Physical Review Letters},
  volume={81},
  number={26},
  pages={5932},
  year={1998},
  publisher={APS},
  doi={10.1103/PhysRevLett.81.5932},
  url={https://doi.org/10.1103/PhysRevLett.81.5932}
}

@article{kimble2008quantum,
  title={The quantum internet},
  author={Kimble, H Jeff},
  journal={Nature},
  volume={453},
  number={7198},
  pages={1023--1030},
  year={2008},
  publisher={Nature Publishing Group},
  doi={10.1038/nature07127},
  url={https://doi.org/10.1038/nature07127}
}

@article{duan2010colloquium,
  title={Colloquium: Quantum networks with trapped ions},
  author={Duan, L-M and Monroe, Christopher},
  journal={Reviews of Modern Physics},
  volume={82},
  number={2},
  pages={1209--1224},
  year={2010},
  publisher={APS},
  doi={10.1103/RevModPhys.82.1209},
  url={https://doi.org/10.1103/RevModPhys.82.1209}
}

@article{wehner2018quantum,
  title={Quantum internet: A vision for the road ahead},
  author={Wehner, Stephanie and Elkouss, David and Hanson, Ronald},
  journal={Science},
  volume={362},
  number={6412},
  pages={eaam9288},
  year={2018},
  publisher={American Association for the Advancement of Science},
  doi={10.1126/science.aam9288},
  url={https://doi.org/10.1126/science.aam9288}
}

@article{wei2022towards,
  title={Towards real-world quantum networks: a review},
  author={Wei, Shi-Hai and Jing, Bo and Zhang, Xue-Ying and Liao, Jin-Yu and Yuan, Chen-Zhi and Fan, Bo-Yu and Lyu, Chen and Zhou, Dian-Li and Wang, You and Deng, Guang-Wei and others},
  journal={Laser \& Photonics Reviews},
  volume={16},
  number={3},
  pages={2100219},
  year={2022},
  publisher={Wiley Online Library},
  doi={10.1002/lpor.202100219},
  url={https://doi.org/10.1002/lpor.202100219}
}

@article{lvovsky2009optical,
  title={Optical quantum memory},
  author={Lvovsky, Alexander I and Sanders, Barry C and Tittel, Wolfgang},
  journal={Nature Photonics},
  volume={3},
  number={12},
  pages={706--714},
  year={2009},
  publisher={Nature Publishing Group UK London},
  doi={10.1038/nphoton.2009.231},
  url={https://doi.org/10.1038/nphoton.2009.231}
}

@article{khabiboulline2019optical,
  title={Optical interferometry with quantum networks},
  author={Khabiboulline, Emil T and Borregaard, Johannes and De Greve, Kristiaan and Lukin, Mikhail D},
  journal={Physical Review Letters},
  volume={123},
  number={7},
  pages={070504},
  year={2019},
  publisher={APS},
  doi={10.1103/PhysRevLett.123.070504},
  url={https://doi.org/10.1103/PhysRevLett.123.070504}
}

@inproceedings{rajagopal2024towards,
  title={Towards quantum-enhanced long-baseline optical/near-IR interferometry},
  author={Rajagopal, Jayadev K and Lau, Ryan M and Padilla, Isack and Ridgway, Stephen T and Cui, Chaohan and McClinton, Brittany and Sajjad, Aqil and Corder, Stuartt and Rawlings, Mark and Rantakyro, Fredrik and others},
  booktitle={Optical and Infrared Interferometry and Imaging IX},
  volume={13095},
  pages={130951N},
  year={2024},
  organization={SPIE},
  doi={10.1117/12.3019518},
  url={https://doi.org/10.1117/12.3019518}
}

@ARTICLE{Padilla2024-in,
  title         = "Quantum resolution limit of long-baseline imaging using
                   distributed entanglement",
  author        = "Padilla, Isack and Sajjad, Aqil and Saif, Babak N and Guha,
                   Saikat",
  journal       = "arXiv [quant-ph]",
  month         =  jun,
  year          =  2024,
  eprint        = "2406.16789",
  archivePrefix = "arXiv",
  primaryClass  = "quant-ph",
  doi={10.48550/arXiv.2406.16789},
  url={https://doi.org/10.48550/arXiv.2406.16789}
}

@ARTICLE{Hermans2023-kf,
  title     = "Entangling remote qubits using the single-photon protocol: an in-depth theoretical and experimental study",
  author    = "Hermans, S L N and Pompili, M and Santos Martins, L Dos and R-P Montblanch, A and Beukers, H K C and Baier, S and Borregaard, J and Hanson, R",
  journal   = "New J. Phys.",
  publisher = "IOP Publishing",
  volume    =  25,
  number    =  1,
  pages     =  013011,
  month     =  jan,
  year      =  2023,
  doi={10.1088/1367-2630/acb004},
  url={https://doi.org/10.1088/1367-2630/acb004}
}

@ARTICLE{Dhara2024-ko,
  title         = "Interfacing Gottesman-Kitaev-Preskill Qubits to Quantum
                   Memories",
  author        = "Dhara, Prajit and Jiang, Liang and Guha, Saikat",
  journal       = "arXiv [quant-ph]",
  month         =  jun,
  year          =  2024,
  eprint        = "2406.04275",
  archivePrefix = "arXiv",
  primaryClass  = "quant-ph",
  doi={10.48550/arXiv.2406.04275},
  url={https://doi.org/10.48550/arXiv.2406.04275}
}

@article{gottesman2012longer,
  title={Longer-baseline telescopes using quantum repeaters},
  author={Gottesman, Daniel and Jennewein, Thomas and Croke, Sarah},
  journal={Physical Review Letters},
  volume={109},
  number={7},
  pages={070503},
  year={2012},
  publisher={APS},
  doi={10.1103/PhysRevLett.109.070503},
  url={https://doi.org/10.1103/PhysRevLett.109.070503}
}

@article{ge2018distributed,
  title={Distributed quantum metrology with linear networks and separable inputs},
  author={Ge, Wenchao and Jacobs, Kurt and Eldredge, Zachary and Gorshkov, Alexey V and Foss-Feig, Michael},
  journal={Physical Review Letters},
  volume={121},
  number={4},
  pages={043604},
  year={2018},
  publisher={APS},
  doi={10.1103/PhysRevLett.121.043604},
  url={https://doi.org/10.1103/PhysRevLett.121.043604}
}

@article{zhang2021distributed,
  title={Distributed quantum sensing},
  author={Zhang, Zheshen and Zhuang, Quntao},
  journal={Quantum Science and Technology},
  volume={6},
  number={4},
  pages={043001},
  year={2021},
  publisher={IOP Publishing},
  doi={10.1088/2058-9565/abd4c3},
  url={https://doi.org/10.1088/2058-9565/abd4c3}
}

@article{jiang2007distributed,
  title={Distributed quantum computation based on small quantum registers},
  author={Jiang, Liang and Taylor, Jacob M and S{\o}rensen, Anders S and Lukin, Mikhail D},
  journal={Physical Review A},
  volume={76},
  number={6},
  pages={062323},
  year={2007},
  publisher={APS},
  doi={10.1103/PhysRevA.76.062323},
  url={https://doi.org/10.1103/PhysRevA.76.062323}
}

@article{monroe2014large,
  title={Large-scale modular quantum-computer architecture with atomic memory and photonic interconnects},
  author={Monroe, Christopher and Raussendorf, Robert and Ruthven, Alex and Brown, Kenneth R and Maunz, Peter and Duan, L-M and Kim, Jungsang},
  journal={Physical Review A},
  volume={89},
  number={2},
  pages={022317},
  year={2014},
  publisher={APS},
  doi={10.1103/PhysRevA.89.022317},
  url={https://doi.org/10.1103/PhysRevA.89.022317}
}

@inproceedings{broadbent2009universal,
  title={Universal blind quantum computation},
  author={Broadbent, Anne and Fitzsimons, Joseph and Kashefi, Elham},
  booktitle={2009 50th annual IEEE symposium on foundations of computer science},
  pages={517--526},
  year={2009},
  organization={IEEE},
  doi={10.1109/FOCS.2009.36},
  url={https://doi.org/10.1109/FOCS.2009.36}
}

@article{barz2012demonstration,
  title={Demonstration of blind quantum computing},
  author={Barz, Stefanie and Kashefi, Elham and Broadbent, Anne and Fitzsimons, Joseph F and Zeilinger, Anton and Walther, Philip},
  journal={Science},
  volume={335},
  number={6066},
  pages={303--308},
  year={2012},
  publisher={American Association for the Advancement of Science},
  doi={10.1126/science.1214707},
  url={https://doi.org/10.1126/science.1214707}
}

@article{moehring2007entanglement,
  title={Entanglement of single-atom quantum bits at a distance},
  author={Moehring, David L and Maunz, Peter and Olmschenk, Steve and Younge, Kelly C and Matsukevich, Dzmitry N and Duan, L-M and Monroe, Christopher},
  journal={Nature},
  volume={449},
  number={7158},
  pages={68--71},
  year={2007},
  publisher={Nature Publishing Group UK London},
  doi={10.1038/nature06118},
  url={https://doi.org/10.1038/nature06118}
}

@article{ritter2012elementary,
  title={An elementary quantum network of single atoms in optical cavities},
  author={Ritter, Stephan and N{\"o}lleke, Christian and Hahn, Carolin and Reiserer, Andreas and Neuzner, Andreas and Uphoff, Manuel and M{\"u}cke, Martin and Figueroa, Eden and Bochmann, Joerg and Rempe, Gerhard},
  journal={Nature},
  volume={484},
  number={7393},
  pages={195--200},
  year={2012},
  publisher={Nature Publishing Group UK London},
  doi={10.1038/nature11023},
  url={https://doi.org/10.1038/nature11023}
}

@article{stephenson2020high,
  title={High-rate, high-fidelity entanglement of qubits across an elementary quantum network},
  author={Stephenson, LJ and Nadlinger, DP and Nichol, BC and An, S and Drmota, P and Ballance, TG and Thirumalai, K and Goodwin, JF and Lucas, DM and Ballance, CJ},
  journal={Physical Review Letters},
  volume={124},
  number={11},
  pages={110501},
  year={2020},
  publisher={APS},
  doi={10.1103/PhysRevLett.124.110501},
  url={https://doi.org/10.1103/PhysRevLett.124.110501}
}

@article{hofmann2012heralded,
  title={Heralded entanglement between widely separated atoms},
  author={Hofmann, Julian and Krug, Michael and Ortegel, Norbert and G{\'e}rard, Lea and Weber, Markus and Rosenfeld, Wenjamin and Weinfurter, Harald},
  journal={Science},
  volume={337},
  number={6090},
  pages={72--75},
  year={2012},
  publisher={American Association for the Advancement of Science},
  doi={10.1126/science.1221856},
  url={https://doi.org/10.1126/science.1221856}
}

@article{knaut2024entanglement,
  title={Entanglement of nanophotonic quantum memory nodes in a telecom network},
  author={Knaut, CM and Suleymanzade, A and Wei, Y-C and Assumpcao, DR and Stas, P-J and Huan, YQ and Machielse, B and Knall, EN and Sutula, M and Baranes, G and others},
  journal={Nature},
  volume={629},
  number={8012},
  pages={573--578},
  year={2024},
  publisher={Nature Publishing Group UK London},
  doi={10.1038/s41586-024-07252-z},
  url={https://doi.org/10.1038/s41586-024-07252-z}
}

@article{van2022entangling,
  title={Entangling single atoms over 33 km telecom fibre},
  author={van Leent, Tim and Bock, Matthias and Fertig, Florian and Garthoff, Robert and Eppelt, Sebastian and Zhou, Yiru and Malik, Pooja and Seubert, Matthias and Bauer, Tobias and Rosenfeld, Wenjamin and others},
  journal={Nature},
  volume={607},
  number={7917},
  pages={69--73},
  year={2022},
  publisher={Nature Publishing Group UK London},
  doi={10.1038/s41586-022-04764-4},
  url={https://doi.org/10.1038/s41586-022-04764-4}
}

@article{bernien2013heralded,
  title={Heralded entanglement between solid-state qubits separated by three metres},
  author={Bernien, Hannes and Hensen, Bas and Pfaff, Wolfgang and Koolstra, Gerwin and Blok, Machiel S and Robledo, Lucio and Taminiau, Tim H and Markham, Matthew and Twitchen, Daniel J and Childress, Lilian and others},
  journal={Nature},
  volume={497},
  number={7447},
  pages={86--90},
  year={2013},
  publisher={Nature Publishing Group UK London},
  doi={10.1038/nature12016},
  url={https://doi.org/10.1038/nature12016}
}

@article{pompili2021realization,
  title={Realization of a multinode quantum network of remote solid-state qubits},
  author={Pompili, Matteo and Hermans, Sophie LN and Baier, Simon and Beukers, Hans KC and Humphreys, Peter C and Schouten, Raymond N and Vermeulen, Raymond FL and Tiggelman, Marijn J and dos Santos Martins, Laura and Dirkse, Bas and others},
  journal={Science},
  volume={372},
  number={6539},
  pages={259--264},
  year={2021},
  publisher={American Association for the Advancement of Science},
  doi={10.1126/science.abg1919},
  url={https://doi.org/10.1126/science.abg1919}
}

@article{awschalom2021development,
  title={Development of quantum interconnects (quics) for next-generation information technologies},
  author={Awschalom, David and Berggren, Karl K and Bernien, Hannes and Bhave, Sunil and Carr, Lincoln D and Davids, Paul and Economou, Sophia E and Englund, Dirk and Faraon, Andrei and Fejer, Martin and others},
  journal={PRX Quantum},
  volume={2},
  number={1},
  pages={017002},
  year={2021},
  publisher={APS},
  doi={10.1103/PRXQuantum.2.017002},
  url={https://doi.org/10.1103/PRXQuantum.2.017002}
}

@article{you2022quantum,
  title={Quantum interference with independent single-photon sources over 300 km fiber},
  author={You, Xiang and Zheng, Ming-Yang and Chen, Si and Liu, Run-Ze and Qin, Jian and Xu, Mo-Chi and Ge, Zheng-Xuan and Chung, Tung-Hsun and Qiao, Yu-Kun and Jiang, Yang-Fan and others},
  journal={Advanced Photonics},
  volume={4},
  number={6},
  pages={066003},
  year={2022},
  publisher={Society of Photo-Optical Instrumentation Engineers},
  doi={10.1117/1.AP.4.6.066003},
  url={https://doi.org/10.1117/1.AP.4.6.066003}
}

@article{dusanowski2022optical,
  title={Optical charge injection and coherent control of a quantum-dot spin-qubit emitting at telecom wavelengths},
  author={Dusanowski, {\L}ukasz and Nawrath, Cornelius and Portalupi, Simone L and Jetter, Michael and Huber, Tobias and Klembt, Sebastian and Michler, Peter and H{\"o}fling, Sven},
  journal={Nature Communications},
  volume={13},
  number={1},
  pages={748},
  year={2022},
  publisher={Nature Publishing Group UK London},
  doi={10.1038/s41467-022-28328-2},
  url={https://doi.org/10.1038/s41467-022-28328-2}
}

@article{yu2023telecom,
  title={Telecom-band quantum dot technologies for long-distance quantum networks},
  author={Yu, Ying and Liu, Shunfa and Lee, Chang-Min and Michler, Peter and Reitzenstein, Stephan and Srinivasan, Kartik and Waks, Edo and Liu, Jin},
  journal={Nature Nanotechnology},
  volume={18},
  number={12},
  pages={1389--1400},
  year={2023},
  publisher={Nature Publishing Group UK London},
  doi={10.1038/s41565-023-01528-7},
  url={https://doi.org/10.1038/s41565-023-01528-7}
}

@article{krutyanskiy2023entanglement,
  title={Entanglement of trapped-ion qubits separated by 230 meters},
  author={Krutyanskiy, V and Galli, M and Krcmarsky, V and Baier, S and Fioretto, DA and Pu, Y and Mazloom, A and Sekatski, P and Canteri, M and Teller, M and others},
  journal={Physical Review Letters},
  volume={130},
  number={5},
  pages={050803},
  year={2023},
  publisher={APS},
  doi={10.1103/PhysRevLett.130.050803},
  url={https://doi.org/10.1103/PhysRevLett.130.050803}
}

@article{stockill2017phase,
  title={Phase-tuned entangled state generation between distant spin qubits},
  author={Stockill, Robert and Stanley, MJ and Huthmacher, Lukas and Clarke, Edmund and Hugues, Maxim and Miller, AJ and Matthiesen, Clemens and Le Gall, Claire and Atat{\"u}re, Mete},
  journal={Physical Review Letters},
  volume={119},
  number={1},
  pages={010503},
  year={2017},
  publisher={APS},
  doi={10.1103/PhysRevLett.119.010503},
  url={https://doi.org/10.1103/PhysRevLett.119.010503}
}

@article{liu2024creation,
  title={Creation of memory--memory entanglement in a metropolitan quantum network},
  author={Liu, Jian-Long and Luo, Xi-Yu and Yu, Yong and Wang, Chao-Yang and Wang, Bin and Hu, Yi and Li, Jun and Zheng, Ming-Yang and Yao, Bo and Yan, Zi and others},
  journal={Nature},
  volume={629},
  number={8012},
  pages={579--585},
  year={2024},
  publisher={Nature Publishing Group UK London},
  doi={10.1038/s41586-024-07308-0},
  url={https://doi.org/10.1038/s41586-024-07308-0}
}

@article{reiserer2015cavity,
  title={Cavity-based quantum networks with single atoms and optical photons},
  author={Reiserer, Andreas and Rempe, Gerhard},
  journal={Reviews of Modern Physics},
  volume={87},
  number={4},
  pages={1379--1418},
  year={2015},
  publisher={APS},
  doi={10.1103/RevModPhys.87.1379},
  url={https://doi.org/10.1103/RevModPhys.87.1379}
}

@article{munro2012quantum,
  title={Quantum communication without the necessity of quantum memories},
  author={Munro, William J and Stephens, Ashley M and Devitt, Simon J and Harrison, Keith A and Nemoto, Kae},
  journal={Nature Photonics},
  volume={6},
  number={11},
  pages={777--781},
  year={2012},
  publisher={Nature Publishing Group UK London},
  doi={10.1038/nphoton.2012.243},
  url={https://doi.org/10.1038/nphoton.2012.243}
}

@article{pant2017rate,
  title={Rate-distance tradeoff and resource costs for all-optical quantum repeaters},
  author={Pant, Mihir and Krovi, Hari and Englund, Dirk and Guha, Saikat},
  journal={Physical Review A},
  volume={95},
  number={1},
  pages={012304},
  year={2017},
  publisher={APS},
  doi={10.1103/PhysRevA.95.012304},
  url={https://doi.org/10.1103/PhysRevA.95.012304}
}

@article{azuma2015all,
  title={All-photonic quantum repeaters},
  author={Azuma, Koji and Tamaki, Kiyoshi and Lo, Hoi-Kwong},
  journal={Nature Communications},
  volume={6},
  number={1},
  pages={6787},
  year={2015},
  publisher={Nature Publishing Group},
  doi={10.1038/ncomms7787},
  url={https://doi.org/10.1038/ncomms7787}
}

@article{lee2022quantum,
  title={A quantum router architecture for high-fidelity entanglement flows in quantum networks},
  author={Lee, Yuan and Bersin, Eric and Dahlberg, Axel and Wehner, Stephanie and Englund, Dirk},
  journal={npj Quantum Information},
  volume={8},
  number={1},
  pages={75},
  year={2022},
  publisher={Nature Publishing Group UK London},
  doi={10.1038/s41534-022-00582-8},
  url={https://doi.org/10.1038/s41534-022-00582-8}
}

@article{pant2019routing,
  title={Routing entanglement in the quantum internet},
  author={Pant, Mihir and Krovi, Hari and Towsley, Don and Tassiulas, Leandros and Jiang, Liang and Basu, Prithwish and Englund, Dirk and Guha, Saikat},
  journal={npj Quantum Information},
  volume={5},
  number={1},
  pages={25},
  year={2019},
  publisher={Nature Publishing Group UK London},
  doi={10.1038/s41534-019-0139-x},
  url={https://doi.org/10.1038/s41534-019-0139-x}
}

@article{sun2016quantum,
  title={A quantum phase switch between a single solid-state spin and a photon},
  author={Sun, Shuo and Kim, Hyochul and Solomon, Glenn S and Waks, Edo},
  journal={Nature Nanotechnology},
  volume={11},
  number={6},
  pages={539--544},
  year={2016},
  publisher={Nature Publishing Group UK London},
  doi={10.1038/nnano.2015.334},
  url={https://doi.org/10.1038/nnano.2015.334}
}

@book{furusawa2011quantum,
  title={Quantum teleportation and entanglement: a hybrid approach to optical quantum information processing},
  author={Furusawa, Akira and Van Loock, Peter},
  year={2011},
  publisher={John Wiley \& Sons},
  doi={10.1002/9783527635283},
  url={https://doi.org/10.1002/9783527635283}
}

@article{hong1987measurement,
  title={Measurement of subpicosecond time intervals between two photons by interference},
  author={Hong, Chong-Ki and Ou, Zhe-Yu and Mandel, Leonard},
  journal={Physical Review Letters},
  volume={59},
  number={18},
  pages={2044},
  year={1987},
  publisher={APS},
  doi={10.1103/PhysRevLett.59.2044},
  url={https://doi.org/10.1103/PhysRevLett.59.2044}
}

@article{dhara2024entangling,
  title={Entangling Quantum Memories at Channel Capacity},
  author={Dhara, Prajit and Jiang, Liang and Guha, Saikat},
  journal={Optica Quantum},
  volume={3},
  pages={480--486},
  year={2025},
  doi={10.1364/OPTICAQ.570931}
}

@article{winnel2022achieving,
  title={Achieving the ultimate end-to-end rates of lossy quantum communication networks},
  author={Winnel, Matthew S and Guanzon, Joshua J and Hosseinidehaj, Nedasadat and Ralph, Timothy C},
  journal={npj Quantum Information},
  volume={8},
  number={1},
  pages={129},
  year={2022},
  publisher={Nature Publishing Group UK London},
  doi={10.1038/s41534-022-00641-0},
  url={https://doi.org/10.1038/s41534-022-00641-0}
}

@article{devetak2005distillation,
  title={Distillation of secret key and entanglement from quantum states},
  author={Devetak, Igor and Winter, Andreas},
  journal={Proceedings of the Royal Society A: Mathematical, Physical and Engineering Sciences},
  volume={461},
  number={2053},
  pages={207--235},
  year={2005},
  publisher={The Royal Society},
  doi={10.1098/rspa.2004.1372},
  url={https://doi.org/10.1098/rspa.2004.1372}
}

@article{vasenin2024evolution,
  title={Evolution of Propagating Coherent Pulses Driving a Single Superconducting Artificial Atom},
  author={Vasenin, AV and Kadyrmetov, Sh V and Bolgar, AN and Dmitriev, A Yu and Astafiev, OV},
  journal={Physical Review Letters},
  volume={133},
  number={7},
  pages={073602},
  year={2024},
  publisher={APS},
  doi={10.1103/PhysRevLett.133.073602},
  url={https://doi.org/10.1103/PhysRevLett.133.073602}
}

@article{sipahigil2016integrated,
  title={An integrated diamond nanophotonics platform for quantum-optical networks},
  author={Sipahigil, Alp and Evans, Ruffin E and Sukachev, Denis D and Burek, Michael J and Borregaard, Johannes and Bhaskar, Mihir K and Nguyen, Christian T and Pacheco, Jose L and Atikian, Haig A and Meuwly, Charles and others},
  journal={Science},
  volume={354},
  number={6314},
  pages={847--850},
  year={2016},
  publisher={American Association for the Advancement of Science},
  doi={10.1126/science.aah6875},
  url={https://doi.org/10.1126/science.aah6875}
}

@article{raymer2024duan,
  title={The Duan-Kimble cavity-atom quantum memory loading scheme revisited},
  author={Raymer, Michael G and Embleton, Clark and Shapiro, Jeffrey H},
  journal={Physical Review Applied},
  volume={22},
  number={4},
  pages={044013},
  year={2024},
  publisher={APS},
  doi={10.1103/PhysRevApplied.22.044013},
  url={https://doi.org/10.1103/PhysRevApplied.22.044013}
}

@article{cui2022quantum,
  title={Quantum receiver enhanced by adaptive learning},
  author={Cui, Chaohan and Horrocks, William and Hao, Shuhong and Guha, Saikat and Peyghambarian, Nasser and Zhuang, Quntao and Zhang, Zheshen},
  journal={Light: Science \& Applications},
  volume={11},
  number={1},
  pages={344},
  year={2022},
  publisher={Nature Publishing Group UK London},
  doi={10.1038/s41377-022-01039-5},
  url={https://doi.org/10.1038/s41377-022-01039-5}
}

@article{cook2007optical,
  title={Optical coherent state discrimination using a closed-loop quantum measurement},
  author={Cook, Robert L and Martin, Paul J and Geremia, John M},
  journal={Nature},
  volume={446},
  number={7137},
  pages={774--777},
  year={2007},
  publisher={Nature Publishing Group UK London},
  doi={10.1038/nature05655},
  url={https://doi.org/10.1038/nature05655}
}

@article{dolinar1973optimum,
  title={An optimum receiver for the binary coherent state quantum channel},
  author={Dolinar, Samuel Joseph},
  journal={Research Laboratory of Electronics, MIT, Quarterly Progress Report},
  volume={11},
  pages={115--120},
  year={1973}
}

@article{duan2004scalable,
  title={Scalable photonic quantum computation through cavity-assisted interactions},
  author={Duan, L-M and Kimble, HJ},
  journal={Physical Review Letters},
  volume={92},
  number={12},
  pages={127902},
  year={2004},
  publisher={APS},
  doi={10.1103/PhysRevLett.92.127902},
  url={https://doi.org/10.1103/PhysRevLett.92.127902}
}

@article{reddy2020superconducting,
  title={Superconducting nanowire single-photon detectors with 98\% system detection efficiency at 1550 nm},
  author={Reddy, Dileep V and Nerem, Robert R and Nam, Sae Woo and Mirin, Richard P and Verma, Varun B},
  journal={Optica},
  volume={7},
  number={12},
  pages={1649--1653},
  year={2020},
  publisher={Optica Publishing Group},
  doi={10.1364/OPTICA.400751},
  url={https://doi.org/10.1364/OPTICA.400751}
}

@article{slussarenko2022quantum,
  title={Quantum channel correction outperforming direct transmission},
  author={Slussarenko, Sergei and Weston, Morgan M and Shalm, Lynden K and Verma, Varun B and Nam, Sae-Woo and Kocsis, Sacha and Ralph, Timothy C and Pryde, Geoff J},
  journal={Nature Communications},
  volume={13},
  number={1},
  pages={1832},
  year={2022},
  publisher={Nature Publishing Group UK London},
  doi={10.1038/s41467-022-29376-4},
  url={https://doi.org/10.1038/s41467-022-29376-4}
}

@article{hilaire2023linear,
  title={Linear optical logical bell state measurements with optimal loss-tolerance threshold},
  author={Hilaire, Paul and Castor, Yaron and Barnes, Edwin and Economou, Sophia E and Grosshans, Fr{\'e}d{\'e}ric},
  journal={PRX Quantum},
  volume={4},
  number={4},
  pages={040322},
  year={2023},
  publisher={APS},
  doi={10.1103/PRXQuantum.4.040322},
  url={https://doi.org/10.1103/PRXQuantum.4.040322}
}

@article{wang2005engineering,
  title={Engineering superpositions of coherent states in coherent optical pulses through cavity-assisted interaction},
  author={Wang, B and Duan, L-M},
  journal={Physical Review A},
  volume={72},
  number={2},
  pages={022320},
  year={2005},
  publisher={APS},
  doi={10.1103/PhysRevA.72.022320},
  url={https://doi.org/10.1103/PhysRevA.72.022320}
}

@article{daiss2019single,
  title={Single-photon distillation via a photonic parity measurement using cavity QED},
  author={Daiss, Severin and Welte, Stephan and Hacker, Bastian and Li, Lin and Rempe, Gerhard},
  journal={Physical Review Letters},
  volume={122},
  number={13},
  pages={133603},
  year={2019},
  publisher={APS},
  doi={10.1103/PhysRevLett.122.133603},
  url={https://doi.org/10.1103/PhysRevLett.122.133603}
}

@article{hacker2019deterministic,
  title={Deterministic creation of entangled atom--light Schr{\"o}dinger-cat states},
  author={Hacker, Bastian and Welte, Stephan and Daiss, Severin and Shaukat, Armin and Ritter, Stephan and Li, Lin and Rempe, Gerhard},
  journal={Nature Photonics},
  volume={13},
  number={2},
  pages={110--115},
  year={2019},
  publisher={Nature Publishing Group UK London},
  doi={10.1038/s41566-018-0339-5},
  url={https://doi.org/10.1038/s41566-018-0339-5}
}

@article{pirandola2009direct,
  title={Direct and reverse secret-key capacities of a quantum channel},
  author={Pirandola, Stefano and Garc{\'\i}a-Patr{\'o}n, Raul and Braunstein, Samuel L and Lloyd, Seth},
  journal={Physical Review Letters},
  volume={102},
  number={5},
  pages={050503},
  year={2009},
  publisher={APS},
  doi={10.1103/PhysRevLett.102.050503},
  url={https://doi.org/10.1103/PhysRevLett.102.050503}
}

@article{pirandola2017fundamental,
  title={Fundamental limits of repeaterless quantum communications},
  author={Pirandola, Stefano and Laurenza, Riccardo and Ottaviani, Carlo and Banchi, Leonardo},
  journal={Nature Communications},
  volume={8},
  number={1},
  pages={15043},
  year={2017},
  publisher={Nature Publishing Group UK London},
  doi={10.1038/ncomms15043},
  url={https://doi.org/10.1038/ncomms15043}
}

@article{dur2007entanglement,
  title={Entanglement purification and quantum error correction},
  author={D{\"u}r, Wolfgang and Briegel, Hans J},
  journal={Reports on Progress in Physics},
  volume={70},
  number={8},
  pages={1381},
  year={2007},
  publisher={IOP Publishing},
  doi={10.1088/0034-4885/70/8/R03},
  url={https://doi.org/10.1088/0034-4885/70/8/R03}
}

@article{waks2006dipole,
  title={Dipole induced transparency in drop-filter cavity-waveguide systems},
  author={Waks, Edo and Vuckovic, Jelena},
  journal={Physical Review Letters},
  volume={96},
  number={15},
  pages={153601},
  year={2006},
  publisher={APS},
  doi={10.1103/PhysRevLett.96.153601},
  url={https://doi.org/10.1103/PhysRevLett.96.153601}
}

@article{dhara2023entangling,
  title={Entangling quantum memories via heralded photonic Bell measurement},
  author={Dhara, Prajit and Englund, Dirk and Guha, Saikat},
  journal={Physical Review Research},
  volume={5},
  number={3},
  pages={033149},
  year={2023},
  publisher={APS},
  doi={10.1103/PhysRevResearch.5.033149},
  url={https://doi.org/10.1103/PhysRevResearch.5.033149}
}

@article{glaudell2016serialized,
  title={Serialized quantum error correction protocol for high-bandwidth quantum repeaters},
  author={Glaudell, Andrew N and Waks, Edo and Taylor, Jacob M},
  journal={New Journal of Physics},
  volume={18},
  number={9},
  pages={093008},
  year={2016},
  publisher={IOP Publishing},
  doi={10.1088/1367-2630/18/9/093008},
  url={https://doi.org/10.1088/1367-2630/18/9/093008}
}

@article{briegel2009measurement,
  title={Measurement-based quantum computation},
  author={Briegel, Hans J and Browne, David E and D{\"u}r, Wolfgang and Raussendorf, Robert and Van den Nest, Maarten},
  journal={Nature Physics},
  volume={5},
  number={1},
  pages={19--26},
  year={2009},
  publisher={Nature Publishing Group UK London},
  doi={10.1038/nphys1157},
  url={https://doi.org/10.1038/nphys1157}
}

@article{lanyon2013measurement,
  title={Measurement-based quantum computation with trapped ions},
  author={Lanyon, BP and Jurcevic, P and Zwerger, M and Hempel, C and Martinez, EA and D{\"u}r, W and Briegel, HJ and Blatt, Rainer and Roos, Christian F},
  journal={Physical Review Letters},
  volume={111},
  number={21},
  pages={210501},
  year={2013},
  publisher={APS},
  doi={10.1103/PhysRevLett.111.210501},
  url={https://doi.org/10.1103/PhysRevLett.111.210501}
}

@article{lucamarini2018overcoming,
  title={Overcoming the rate--distance limit of quantum key distribution without quantum repeaters},
  author={Lucamarini, Marco and Yuan, Zhiliang L and Dynes, James F and Shields, Andrew J},
  journal={Nature},
  volume={557},
  number={7705},
  pages={400--403},
  year={2018},
  publisher={Nature Publishing Group UK London},
  doi={10.1038/s41586-018-0066-6},
  url={https://doi.org/10.1038/s41586-018-0066-6}
}

@article{wang2019beating,
  title={Beating the fundamental rate-distance limit in a proof-of-principle quantum key distribution system},
  author={Wang, Shuang and He, De-Yong and Yin, Zhen-Qiang and Lu, Feng-Yu and Cui, Chao-Han and Chen, Wei and Zhou, Zheng and Guo, Guang-Can and Han, Zheng-Fu},
  journal={Physical Review X},
  volume={9},
  number={2},
  pages={021046},
  year={2019},
  publisher={APS},
  doi={10.1103/PhysRevX.9.021046},
  url={https://doi.org/10.1103/PhysRevX.9.021046}
}

@article{bersin2024telecom,
  title={Telecom networking with a diamond quantum memory},
  author={Bersin, Eric and Sutula, Madison and Huan, Yan Qi and Suleymanzade, Aziza and Assumpcao, Daniel R and Wei, Yan-Cheng and Stas, Pieter-Jan and Knaut, Can M and Knall, Erik N and Langrock, Carsten and others},
  journal={PRX Quantum},
  volume={5},
  number={1},
  pages={010303},
  year={2024},
  publisher={APS},
  doi={10.1103/PRXQuantum.5.010303},
  url={https://doi.org/10.1103/PRXQuantum.5.010303}
}

@article{sidhu2023linear,
  title={Linear optics and photodetection achieve near-optimal unambiguous coherent state discrimination},
  author={Sidhu, Jasminder S and Bullock, Michael S and Guha, Saikat and Lupo, Cosmo},
  journal={Quantum},
  volume={7},
  pages={1025},
  year={2023},
  publisher={Verein zur F{\"o}rderung des Open Access Publizierens in den Quantenwissenschaften},
  doi={10.22331/q-2023-05-31-1025},
  url={https://doi.org/10.22331/q-2023-05-31-1025}
}

@article{duvsek2000unambiguous,
  title={Unambiguous state discrimination in quantum cryptography with weak coherent states},
  author={Du{\v{s}}ek, Miloslav and Jahma, Mika and L{\"u}tkenhaus, Norbert},
  journal={Physical Review A},
  volume={62},
  number={2},
  pages={022306},
  year={2000},
  publisher={APS},
  doi={10.1103/PhysRevA.62.022306},
  url={https://doi.org/10.1103/PhysRevA.62.022306}
}

@article{mohseni2004optical,
  title={Optical Realization of Optimal Unambiguous Discrimination for Pure and Mixed Quantum States},
  author={Mohseni, Masoud and Steinberg, Aephraim M and Bergou, J{\'a}nos A},
  journal={Physical Review Letters},
  volume={93},
  number={20},
  pages={200403},
  year={2004},
  publisher={APS},
  doi={10.1103/PhysRevLett.93.200403},
  url={https://doi.org/10.1103/PhysRevLett.93.200403}
}

@article{vardoyan2020exact,
  title={On the exact analysis of an idealized quantum switch},
  author={Vardoyan, Gayane and Guha, Saikat and Nain, Philippe and Towsley, Don},
  journal={Performance Evaluation},
  volume={144},
  pages={102141},
  year={2020},
  publisher={Elsevier},
  doi={10.1016/j.peva.2020.102141},
  url={https://doi.org/10.1016/j.peva.2020.102141}
}

@article{bhaskar2020experimental,
  title={Experimental demonstration of memory-enhanced quantum communication},
  author={Bhaskar, Mihir K and Riedinger, Ralf and Machielse, Bartholomeus and Levonian, David S and Nguyen, Christian T and Knall, Erik N and Park, Hongkun and Englund, Dirk and Lon{\v{c}}ar, Marko and Sukachev, Denis D and others},
  journal={Nature},
  volume={580},
  number={7801},
  pages={60--64},
  year={2020},
  publisher={Nature Publishing Group UK London},
  doi={10.1038/s41586-020-2103-5},
  url={https://doi.org/10.1038/s41586-020-2103-5}
}

@article{delteil2016generation,
  title={Generation of heralded entanglement between distant hole spins},
  author={Delteil, Aymeric and Sun, Zhe and Gao, Wei-bo and Togan, Emre and Faelt, Stefan and Imamo{\u{g}}lu, Ata{\c{c}}},
  journal={Nature Physics},
  volume={12},
  number={3},
  pages={218--223},
  year={2016},
  publisher={Nature Publishing Group UK London},
  doi={10.1038/nphys3605},
  url={https://doi.org/10.1038/nphys3605}
}

@article{konno2024logical,
  title={Logical states for fault-tolerant quantum computation with propagating light},
  author={Konno, Shunya and Asavanant, Warit and Hanamura, Fumiya and Nagayoshi, Hironari and Fukui, Kosuke and Sakaguchi, Atsushi and Ide, Ryuhoh and China, Fumihiro and Yabuno, Masahiro and Miki, Shigehito and others},
  journal={Science},
  volume={383},
  number={6680},
  pages={289--293},
  year={2024},
  publisher={American Association for the Advancement of Science},
  doi={10.1126/science.adk7560},
  url={https://doi.org/10.1126/science.adk7560}
}

@article{shi2025measurement,
  title     = {Measurement-Based Entanglement Distillation and Constant-Rate Quantum Repeaters over Arbitrary Distances},
  author    = {Shi, Yu and Patil, Ashlesha and Guha, Saikat},
  journal   = {Physical Review Letters},
  volume    = {135},
  number    = {13},
  pages     = {130803},
  year      = {2025},
  month     = sep,
  publisher = {American Physical Society},
  doi       = {10.1103/2bp8-cdxc},
  url       = {https://doi.org/10.1103/2bp8-cdxc}
}

@article{pant2019percolation,
  title={Percolation thresholds for photonic quantum computing},
  author={Pant, Mihir and Towsley, Don and Englund, Dirk and Guha, Saikat},
  journal={Nature Communications},
  volume={10},
  number={1},
  pages={1070},
  year={2019},
  publisher={Nature Publishing Group UK London},
  doi={10.1038/s41467-019-08948-x},
  url={https://doi.org/10.1038/s41467-019-08948-x}
}

@article{patil2026clifford,
  title     = {{Clifford} Manipulations of Stabilizer States--Application to Linear Optical Qubits},
  author    = {Patil, Ashlesha and Guha, Saikat},
  journal   = {IEEE Transactions on Quantum Engineering},
  year      = {2026},
  month     = aug,
  pages     = {1--16},
  doi       = {10.1109/TQE.2026.3721271},
  issn      = {2689-1808},
  publisher = {IEEE},
  url = {https://ieeexplore.ieee.org/abstract/document/11643553}
}

@article{ewert2014efficient,
  title     = {$3/4$-Efficient {Bell} Measurement with Passive Linear Optics and Unentangled Ancillae},
  author    = {Ewert, Fabian and van Loock, Peter},
  journal   = {Physical Review Letters},
  volume    = {113},
  number    = {14},
  pages     = {140403},
  year      = {2014},
  month     = sep,
  publisher = {American Physical Society},
  doi       = {10.1103/PhysRevLett.113.140403},
  url       = {https://doi.org/10.1103/PhysRevLett.113.140403}
}

\end{document}